\documentclass{aa}

\usepackage{graphicx}
	\graphicspath{{./img/},{./cartoons/},{./}}
\usepackage{txfonts}
\usepackage{hyperref}
\usepackage{xcolor}
\usepackage{textcomp,mathabx}
\usepackage{siunitx}
    \DeclareSIUnit{\astronomicalunit}{AU}
	\DeclareSIUnit{\parsec}{pc}
	\DeclareSIUnit{\earthmass}{M_\Earth}
	\DeclareSIUnit{\solarmass}{M_\Sun}
	\DeclareSIUnit{\jupitermass}{M_J}
	\DeclareSIUnit{\year}{yr}

\begin{document} 

    \bibliographystyle{aa}
    \title{How does accretion of planet-forming disks influence stellar abundances?}

    \author{L.-A. H\"uhn\inst{1,2}
        \and
        B. Bitsch\inst{2}
        }

    \institute{Zentrum für Astronomie der Universität Heidelberg, %
            Institut für Theoretische Astrophysik, Albert-Ueberle-Str. 2, %
            69120 Heidelberg\\
            \email{huehn@uni-heidelberg.de}
        \and
            Max Planck Institute for Astronomy, Königstuhl 17, 69117 Heidelberg, Germany
            }

    \date{\today}
 
    \abstract
    {Millimeter sized dust grains experience radial velocities exceeding the gas velocities by orders of magnitude. The viscous evolution of the accretion disk adds disk material onto the central star's convective envelope, influencing its elemental abundances, [X/H]. At the same time, the envelope mass shrinks as the stellar age increases, amplifying the rate of abundance change. Therefore, the elemental abundances of the star are sensitive to disk processes that alter the composition and timing of disk accretion. We perform numerical 1D log-radial simulations integrating the disk advection-diffusion equation, while accounting for evaporation and condensation of chemical species at the evaporation fronts. They reveal a peak of refractory abundance within the first \SI{2}{\mega\year} of $\Delta\mathrm{[X/H]}\sim\num{5e-2}$ if grain growth is significant, but subsequent accretion diminishes previous refractory abundance increases for long-lived disks. Planet formation can reduce the abundance of dust species whose evaporation fronts lie within the planet's orbit by preventing solids from reaching the inner edge once the planet starts opening a gap exerting a pressure bump exterior to its orbit and consequently blocking inward drifting pebbles. We expect the accretion of the Solar protoplanetary disk with Jupiter present to have changed the Sun's elemental abundances by ${\sim}\num{1e-2}$ throughout its lifetime. These considerations are also applied to the HD106515 wide binary system. We find that measurements of $\Delta\mathrm{[X/H]}$ are in reasonable agreement with results from simulations where the observed giant planet around HD106515 A is included and if HD106515B's disk formed planetesimals more efficiently. Simulations where the planet formed inside the water ice line are more favorable to agree with observations. Even though the general changes in the stellar abundances due to disk accretion are small, they are detectable at current sensitivities, indicating that the here presented methods can be used to constrain the planet formation pathway.}


    \maketitle
%
\section{Introduction}
The formation of exoplanets takes place in protoplanetary disks around young host stars, consisting of mainly hydrogen and helium gas, but also heavier elements in both solid and gaseous form. Their presence is a natural outcome of star formation (for a review, see \citealt{williams2011}). In these disks, planet cores grow by accreting material from the disk. As this process takes place around the young host star, it is apparent that the stellar evolution cannot be treated as taking place in an isolated system. While the stellar irradiation is a common aspect considered in planet formation models as a form of stellar influence on the disk (e.g., \citealt{chiang1997,dullemond2004,bitsch2015a,savvidou2020}), the reverse impact of the surroundings on the star is not to be neglected.

The large fraction of stars hosting at least one planet naturally leads to the conclusion that planet formation is a ubiquitous phenomenon, further arguing for its consideration in the study of young stars. The protostar and its disk initially share the same chemical composition, having formed from collapsing molecular cloud material. With the evolution of the protoplanetary disk, possibly including planet formation, and the rapid dispersal of the disk at the end of its lifetime due to photoevaporation (for a review, see \citealt{alexander2014}), the distribution of material in the disk may change. In turn, this may lead to the star changing its chemical composition as it accretes disk material. For example, neglecting the effects of planet formation, material accreted onto the star is initially enriched in mm-cm sized pebbles, because they experience radial drift and have a radial velocity that exceeds that of the gas \citep{weidenschilling1977,brauer2008}. The accreted enriched material cannot be diluted by the depleted gas remaining in the disk, as it is dispersed before the entire disk mass can be accreted. Planet formation can then influence the accretion of these solids, either through taking away material from the disk that can then not be accreted onto the star, or by exerting pressure bumps in the disc that block inward drifting pebbles. The diversity of the exoplanet sample, which in turn raises ideas of different formation scenarios and different formation locations of planets, could mean different abundance changes of the host star depending on the planetary system they harbor.

Unlike for exoplanet atmospheres, where observations of atmospheric chemical abundances prove to be challenging at this time, stellar spectra and therefore elemental abundances can be readily observed. With the above concepts in mind, observations of young stars could provide indirect information about the protoplanetary disks they host, and formed or forming planets therein. Previous studies have employed such an approach to study the depletion of refractory elements in the Sun compared to Solar twins \citep{melendez2009} by considering relative differences in the accretion rate of gas and dust (e.g., \citealt{hoppe2020}). Such a difference was previously considered as a result of locking up solids in rocky planets \citep{chambers2010}, but may more likely be explained by pressure bumps created as a result of giant planet formation \citep{booth2020}. The latter idea is supported by observations of transition disks, which is a group spanning about 10-20\% of protoplanetary disks characterized by large gaps (e.g., \citealt{kenyon1995, koepferl2013, vdmarel2018}), which could be linked to planets (e.g., \citealt{owen2012}). Additionally, the observed gaps and rings in mm observations with ALMA (e.g., \citealt{andrews2018}) are also caused by pressure perturbations preventing a smooth inward drift of pebbles (e.g., \citealt{pinilla2012}). The link between gaps and stellar abundance differences was previously explored for young A stars, which offer a direct insight into the recent accretion history due to their small convective zones \citep{jermyn2018}. This also follows from the correlation between stellar mass and convective zone size during pre-MS evolution, as shown in Fig. \ref{fig:conv_zone_time}. It was shown that there is a correlation between large disk cavities and depletion of refractory elements \citep{kama2015}, which can be observed as a prominent feature due to the absence of convective mixing at the investigated stellar masses. For solar- and sub-solar-mass stars, convective mixing reduces the impact of accreted material, but studies investigating the composition of gas accreting onto T Tauri stars and TW Hya directly have shown a similar link \citep{mcclure2019,mcclure2020}. 

The same concept can also be applied to stellar binary systems. Stellar binaries with wide separations (from ${\sim}\SI{100}{\astronomicalunit}$ to ${\sim}\SI{1}{\parsec}$) form from the same molecular cloud at the same time \citep{kouwenhoven2010,reipurth2012} and are therefore expected to exhibit the same chemical abundances. Observations show, however, that this is not always the case \citep{teske2016b,saffe2019,liu2021}. \citet{liu2021} used high quality spectra from VLT/UVES and Keck/HIRES to derive detailed abundance differences of a wide variety of chemical elements for several wide binary systems containing a different number of planets. The abundance differences could be linked to the presence of a planet around one star, but not the other, in particular for HD106515, where only one constituent harbors exactly one planet \citep{li2021}. Since close binaries would exhibit each other's disk evolution, e.g., by stellar radiation, such studies are limited to wide binaries where these effects are negligible. Previously, differences in chemical abundances in wide binaries were used to investigate the formation location of giant planets with respect to the water ice line, under the premise that locking up material in those planets is the dominant effect (e.g., \citealt{tuccimaia2014, ramirez2015, teske2016a, bitsch2018b}). This approach was also applied to the formation location of super-Earths with respect to the CO ice line \citep{bitsch2018b}.

\begin{figure*}[htp]
    \centering\includegraphics[width=\textwidth]{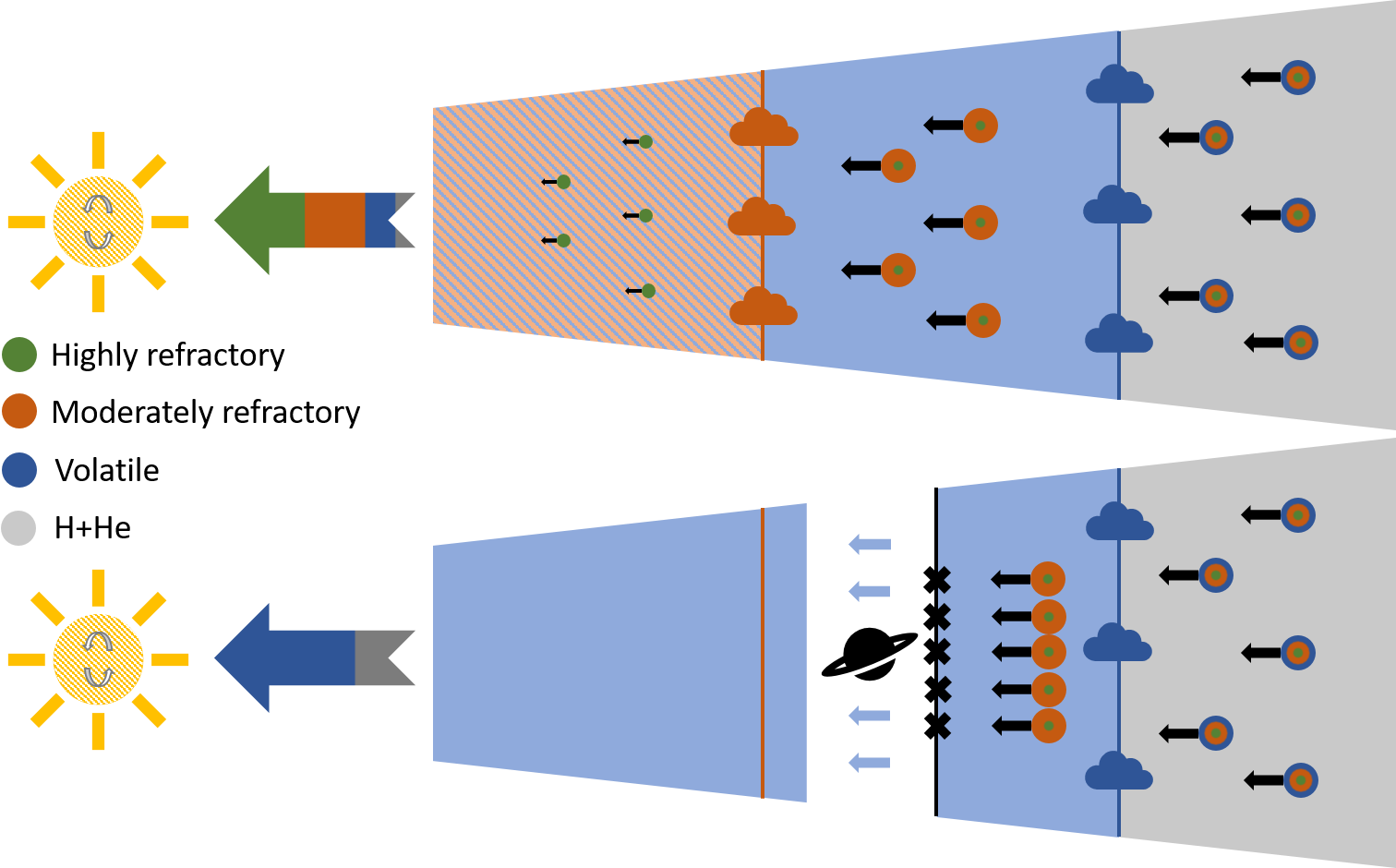}
    \caption{Qualitative illustration of the impact of dust drift and planet formation on the composition of material accreted by the host star. The top part shows the drift of pebbles consisting of three different chemical species, a volatile one (blue), a moderately refractory one (orange) and a highly refractory one (green). The ice lines of the blue and orange species are shown with vertical lines of the corresponding color, while the background color of the disk indicates the gas components. The green species does not evaporate in the disk. More refractory species drift further before evaporating at the ice line, enriching the accreted material more strongly than volatile species. The bottom illustration shows how a giant planet core at pebble isolation mass changes the accreted material's composition. The volatile component evaporates before the pebbles reach the planet's pressure bump, allowing diffusion past the planet's orbit and onto the star. The orange and green species are solid at the pressure bump location and blocked from further accretion. Therefore, the accreted material is no longer enriched in the moderately (orange) and highly refractory (green) species.}
    \label{fig:model_cartoon}
\end{figure*}
In this work, the convective envelope elemental abundance change after the accretion of disk material is obtained and compared to the initial condition. In a parameter study, we aim to give a perspective on initial conditions commonly applied for planet formation models. Furthermore, the investigation of the influence of planets on stellar abundances is continued, with a particular focus on the wide binary system HD106515. However, instead of considering locking up material in the planets as the main mechanism, pressure bumps acting as traps for dust, but not for gas are considered \citep{rice2006, pinilla2012, zhu2014}. These bumps can be created by giant planets reaching a limiting mass to carve a partial gap into the disk's surface density, the so-called pebble-isolation mass \citep{morbidelli2012,lambrechts2014,bitsch2018a,ataiee2018,paardekooper2006}, typically also opening a deep gap into the disk in subsequent evolution \citep{lin1986, crida2006}. The approach is in principle similar to previous works on the elemental abundances of the Sun \citep{booth2020,hoppe2020}. Key differences lie in the differentiation of solid material accreted onto the star into specific chemical species, as well as a more detailed treatment of planetary growth by pebble accretion. The basic processes that lie the foundation for this work are shown conceptually in Fig. \ref{fig:model_cartoon}. The model applied will employ a chemical partitioning model to distribute the various chemical elements into molecules and track their movement through the disk as part of the overall viscous evolution in both gaseous and solid phases.

In Sect. \ref{sec:methods}, we will give a brief overview over some core models implemented in the \texttt{chemcomp} code designed to represent the processes described above. Section \ref{sec:results} is used to present the results of the study, where the change to convective zone abundances due to disk accretion are considered. First, we perform a parameter study for cases without a growing giant planet, discussing viscosity and fragmentation velocity (Sect. \ref{sec:res_nopl}), initial stellar mass and metallicity (Sect. \ref{sec:res_nopl_stellar}) and the impact of planetesimal formation (Sect. \ref{sec:res_nopl_planetesimal}). After that, results from simulations including a giant planet are presented (Sect. \ref{sec:res_pl}), where the influence of the planet's formation location (Sect \ref{sec:res_pl_loc}), as well as implications for Jupiter in the Solar disk (Sect. \ref{sec:res_jup}) are discussed. In Sec. \ref{sec:res_hd}, we present the application to the HD106515 binary system. Finally, Sect. \ref{sec:discussion} will discuss the results and Sect. \ref{sec:summary} will give a brief summary.

\section{Methods}\label{sec:methods}
We use the \texttt{chemcomp} code \citep{schneider2021} to simulate the diffusive evolution of the disk, including radial drift of the dust. To be able to find the individual elemental abundances in the convective zone of the host star, the chemical composition in both gas and dust in the disk is tracked. In addition, the code includes a model for the growth of a giant planet opening a gap in the disk. In this section, the core aspects of these models will be highlighted. More details can be found in \citet{schneider2021}.

\subsection{Dust drift}\label{sec:dust_evolution}
As the main ingredient for the time evolution of the disk's solid component, the two-component model by \citet{birnstiel2012} is used. Based on numerical and analytical considerations, this model simplifies the treatment of the grain growth and size distribution to save computation time by only treating a population of the smallest grains and one of grains close to size limit set by fragmentation and drift. A more detailed description of the applied dust drift model is given in Appendix \ref{sec:dust_drift}.

The dust velocity is set by representative sizes of the two populations and is evolved with a single advection-diffusion equation,
\begin{equation}
    \frac{\partial\Sigma_Z}{\partial t} + \frac{1}{r}\frac{\partial}{\partial r}\left[r\left(\Sigma_Z\bar{u}_Z-\nu\Sigma_\mathrm{gas}\frac{\partial}{\partial r}\left(\frac{\Sigma_Z}{\Sigma_\mathrm{gas}}\right)\right)\right] = -S-S_\mathrm{acc}-S_\mathrm{pla}\text{,}\label{eq:adv_diff_dust}
\end{equation}
where $\nu$ is the kinematic viscosity and $\bar{u}_Z$ is the mass-weighted velocity,
\begin{equation}
    \bar{u}_Z = (1-f_m)u_\mathrm{small}+f_m u_\mathrm{large}\text{,}\label{eq:uz}
\end{equation}
with $u_\mathrm{small}$ and $u_\mathrm{large}$ the radial drift velocities for the representative sizes of the smaller and large population, respectively, as described by Eq. \eqref{eq:driftspeed}. The model parameter $f_m$ describes the mass distribution of the populations, i.e. $\Sigma_\mathrm{large} = \Sigma_Z f_m$, implemented as $f_m = 0.97$ in the drift limited case and $f_m=0.75$ in the fragmentation limited case. It is imperative to note here that, despite the original division of the dust into two populations with distinct size and resulting velocity, the dust component as a whole is advected with a singular, mass-averaged, velocity, as given by Eqs. \eqref{eq:uz} and \eqref{eq:adv_diff_dust}. For the dynamics, that is a reasonable assumption, as the large grain population accounts for 95\% or 75\% of the dust mass in the drift and fragmentation limited regime, respectively. However, this approximation prevents a treatment of small grains being accreted through the planetary pressure bump (\citealt{stammler2023,ataiee2018,bitsch2018a}; see also Sect. \ref{sec:disc_grainsize}).

The right-hand side of Eq. \eqref{eq:adv_diff_dust} contains the source term describing evaporation of dust, $S$ (see Sect. \ref{sec:meth_chem}), and the sink term $S_\mathrm{acc}$ related to pebble accretion of a potential giant planet. In addition, it contains the sink term related to planetesimal formation $S_\mathrm{pla}$, discussed in the following.
\subsection{Planetesimal formation}
In the \texttt{chemcomp} code, a pebble flux-regulated planetesimal formation model is used \citep{lenz2019,voelkel2020}, which assumes a continuous mechanism that converts a fraction of the local pebble flux into planetesimals, preventing that material from partaking in the enrichment of the central star's convective envelope. The sink term $S_\mathrm{pla}$ in Eq. \eqref{eq:adv_diff_dust} is given by
\begin{equation}
    S_\mathrm{pla} = \dot{\Sigma}_\mathrm{pla} = \frac{\epsilon}{d(r)}\frac{\dot{M}_\mathrm{peb}}{2\pi r}\text{.}\label{eq:sigdot_pla}
\end{equation}
Hence, planetesimal formation can prevent heavy elements from being accreted onto the central star by locking up solids as they are created, because they are decoupled from the protoplanetary disk and do not drift inward. The sink term in Eq. \eqref{eq:sigdot_pla} includes the pebble flux $\dot{M}_\mathrm{peb}$, the radial separation of pebble traps,
\begin{equation}
    d(r) = d_\mathrm{pla}H_\mathrm{gas}\text{,}
\end{equation}
and the efficiency parameter $\epsilon$, which describes the fraction of the pebble flux used for the formation of planetesimals after drifting the distance $d$. Both the separation in terms of the gas pressure scale height, $d_\mathrm{pla}$, and efficiency $\epsilon$ are model parameters. The former is kept constant at $d_\mathrm{pla}=5$, the latter at $\epsilon=0.05$. The efficiency is varied in Sect. \ref{sec:res_hd}. Different from the original model by \citet{lenz2019}, only the large grain population is considered for the pebble flux, but there is no specific Stokes number cutoff. Furthermore, a threshold pebble flux must be reached, which is set by the condition that the mass fraction which can be converted at a given location is high enough such that at least one planetesimal can be formed,
\begin{equation}
    \dot{M}_\mathrm{crit} = \frac{M_\mathrm{pla}}{\epsilon\tau_t}\text{,}\label{eq:mdot_pla_crit}
\end{equation}
where $M_\mathrm{pla}=\frac{4}{3}\pi \rho_\mathrm{pla}R^3_\mathrm{pla}$ is the mass of one planetesimal, with $\rho_\mathrm{pla}=\SI{1}{\gram\per\cubic\centi\meter}$ and $R_\mathrm{pla}=\SI{50}{\kilo\meter}$ kept constant, corresponding to the characteristic size of planetesimals formed by the streaming instability \citep{johansen2015,simon2015,klahr2020}, and the trap lifetime,
\begin{equation}
    \tau_t = 100\times\frac{2\pi}{\Omega_K}\text{.}
\end{equation}
At disk locations where the critical flux is not reached, the sink term is identically zero.

\subsection{Disk chemical composition}\label{sec:meth_chem}
Initially, the chemical composition of the protoplanetary disk is assumed to be identical to the composition of the star, which is motivated by the disk formation process, where both are made of the same molecular cloud material. The initial composition is defined by the initial [X/H] values defined relative to the Solar measurements \citep{asplund2009}. Together with the solid-to-gas ratio $\epsilon_d$, taken at a temperature where all available solids are condensed ($T<\SI{20}{\kelvin}$), the absolute mass of every chemical element is set.

\begin{table*}[htp]
    \centering
    \caption[Volume mixing rations of partitioning model]{Volume mixing ratios for the considered molecular species, given by the applied chemical partitioning model, along with the condensation temperatures of the respective species.}
    \vspace*{1em}
    \begin{tabular}{ccc}
    \hline
    \hline
        Y & $T_\mathrm{cond}$ [K] & $v_\text{Y}$\\
    \hline
        CO & 20 & 0.2 $\times$ C/H\\
        N$_2$ & 20 & 0.45 $\times$ N/H\\
        CH$_4$ & 30 & 0.1 $\times$ C/H\\
        CO$_2$ & 70 & 0.1 $\times$ C/H\\
        NH$_3$ & 90 & 0.1 $\times$ N/H\\
        H$_2$S & 150 & 0.1 $\times$ S/H\\
        H$_2$O & 150 & \multicolumn{1}{m{21em}}{O/H - (3 $\times$ MgSiO$_3$/H + 4 $\times$ Mg$_2$SiO$_4$/H \newline+ CO/H + 2 $\times$ CO$_2$/H + 3 $\times$ Fe$_2$O$_3$/H \newline+ 4 $\times$ Fe$_3$O$_4$/H + VO/H + TiO/H \newline+ 3 $\times$ Al$_2$O$_3$/H + 8 $\times$ NaAlSi$_3$O$_8$/H \newline+ 8 $\times$ KAlSi$_3$O$_8$/H)}\\
        Fe$_3$O$_4$ & 371 & (1/6) $\times$ (Fe/H - 0.9 $\times$ S/H)\\
        C & 631 & 0.6 $\times$ C/H\\
        FeS & 704 & 0.9 $\times$ S/H\\
        NaAlSi$_3$O$_8$ & 958 & Na/H\\
        KAlSi$_3$O$_8$ & 1006 & K/H\\
        Mg$_2$SiO$_4$ & 1354 & Mg/H - (Si/H - 3 $\times$ K/H - 3 $\times$ Na/H)\\
        Fe$_2$O$_3$ & 1357 & 0.25 $\times$ (Fe/H - 0.9 $\times$ S/H)\\
        VO & 1423 & V/H\\
        MgSiO$_3$ & 1500 & Mg/H - 2 $\times$ (Mg/H - (Si/H - 3 $\times$ K/H - 3 $\times$ Na/H))\\
        Al$_2$O$_3$ & 1653 & 0.5 $\times$ (Al/H - (K/H + Na/H))\\
        TiO & 2000 & Ti/H\\
    \hline
    \end{tabular}
    \label{tab:partitioning}
\end{table*}
The chemical elements are distributed into molecular species Y according to the chemical partitioning model described in \citet{schneider2021}. This simple model does not treat changes in the solid composition caused, e.g., by chemical reactions, but tracks the relevant main carriers. This approach is justified, because normally the chemical reaction time scales are longer than the drift timescales, leaving inward drifting pebbles unaltered \citep{booth2019,eistrup2022}. Table \ref{tab:partitioning} provides the corresponding volume mixing ratios based on the disk elemental abundances, along with the condensation temperatures of the molecules \citep{lodders2003}. Initially, the entire mass of molecule Y will be in the gas phase in the interior disk region where the temperature is larger than the condensation temperature, and completely in the solid phase exterior to that. During the simulation, transitions between solid and gas phases of the dust species in both directions are possible, accounting for evaporation as molecules cross their ice lines from colder to warmer areas in the disk, as well as for condensation in the opposite case\footnote{To guarantee mass conservation, only a maximum of 90\% of the mass of a given element and at a given location is allowed to change phase per time step $\Delta t$.}. This is reflected in the source term $S$ (Eq. \ref{eq:adv_diff_dust}) and a corresponding term for the gas diffusion equation, given for species Y as
\begin{equation}
    S_Y = \dot{\Sigma}_Y = \left\{\begin{matrix}\dot{\Sigma}_Y^\mathrm{evap} & r < r_\mathrm{ice,Y}\text{,}\\ \dot{\Sigma}_Y^\mathrm{cond} & r \geq r_\mathrm{ice,Y}\text{,}\end{matrix}\right.
\end{equation}
where $r_\mathrm{ice,Y}$ describes the radial coordinate where $T(r_\mathrm{ice,Y})=T_\mathrm{cond,Y}$ in the disk. Under the assumption that dust can only condensate onto pre-existing grains, the condensation term reads
\begin{equation}
    \dot{\Sigma}_Y^\mathrm{cond} = \frac{3\epsilon_p}{2\pi\rho_s}\Sigma_\mathrm{gas,Y}\left(\frac{\Sigma_Z}{a_\mathrm{small}}+\frac{\Sigma_\mathrm{large}}{a_\mathrm{large}}\right)\Omega_K\sqrt{\frac{\mu}{\mu_Y}}\text{,}\label{eq:sigdot_cond}
\end{equation}
with $\mu$ the mean molecular weight of the disk, $\mu_Y$ the mass of the molecular species Y, $a$ the representative size of the small and large dust component, respectively, $\Omega_K$ the Keplerian angular frequency, and $\epsilon_p=0.5$ the sticking efficiency. The evaporation term reads
\begin{equation}
    \dot{\Sigma}_Y^\mathrm{evap} = \frac{\Sigma_{Z,Y}\bar{u}_Z}{\SI{1e-3}{\astronomicalunit}}\text{,}
\end{equation}
where it is assumed that solids passing their ice line evaporate within $\SI{1e-3}{\astronomicalunit}$.

The integration of Eq. \ref{eq:adv_diff_dust} and the analogous equation for the gas is done using a modified version of the flux-conserving donor-cell scheme \citep{birnstiel2010}. The particular implementation utilized in \texttt{chemcomp} is adapted from the unpublished \texttt{DISKLAB} code \citep{dullemond2018}. The disk radial grid extends from \SI{0.1}{\astronomicalunit} to \SI{1000}{\astronomicalunit} and is divided into 500 cells whose sizes are distributed logarithmically.

\subsection{Convective zone abundances}
Solid and gaseous disk material crossing the inner edge is accreted onto the convective zone of the star and mixed with the material therein. To investigate this, we added routines to \texttt{chemcomp} that calculate the change of the convective zone abundances based on the flux onto the star and its mass. To obtain the mass of material that is accreted, the flux of gas and solids at the inner edge is integrated. The gas flux is given by
\begin{equation}
    \dot{M}_\mathrm{gas} = -2\pi r\frac{3}{\sqrt{r}}\frac{\partial}{\partial r}\left(\nu\Sigma_\mathrm{gas}\sqrt{r}\right)\text{.}
\end{equation}
As seen in Eq. \ref{eq:adv_diff_dust}, the mass flux for solids reads
\begin{equation}
    \dot{M}_\mathrm{dust} = 2\pi r\Sigma_Z\bar{u}_Z-2\pi r\Sigma_\mathrm{gas}\nu\frac{\partial}{\partial r}\left(\frac{\Sigma_Z}{\Sigma_\mathrm{gas}}\right)\text{.}\label{eq:mdot}
\end{equation}
The abundance of the convective envelope is advanced in time as follows. First, the mass that was accreted during a given time step from $t_k$ to $t_{k+1}$ is added to the envelope. In detail, the elemental abundances and total mass of the envelope at time $t_k$ are used to calculate the total mass per element at time $t_k$, $M_{i,k}$,
\begin{equation}
    M_{i,k} = \frac{M_{\mathrm{CE},k}\left(\frac{N_i}{N_H}\right)_\odot\frac{m_i}{m_H}\times 10^{[i/H]}}{1+\sum\limits_{j \in \{\mathrm{X}\}\setminus \{\mathrm{H}\}}\left(\frac{N_j}{N_H}\right)_\odot\frac{m_j}{m_H}\times 10^{[j/H]}}\text{,}
\end{equation}
where the sum is taken over all modeled elements except hydrogen, and $m_i$ indicates the atomic mass of element $i$. This approach holds true under the assumption that the envelope preserves its composition during its evolution. After adding the accreted mass,
\begin{equation}
    M_{i,k+1} = M_{i,k} + \int_{t_k}^{t_{k+1}}\left(\dot{M}_\mathrm{gas,k}+\dot{M}_\mathrm{dust,k}\right)dt\text{,}
\end{equation}
the elemental abundances are recalculated. This implies that the mixing is instantaneous. During the next time step, the procedure is repeated for a new convective envelope mass $M_{\mathrm{CE},k+1}$ as given by the convective zone evolution. Overall, material accreted at a time when the convective zone is massive has a smaller impact on its abundance. Therefore, as material with a particular [X/H] is accreted, it takes longer for the abundances in the convective zone to approach that value if the convective zone is massive than if it is light.

\begin{figure*}[htp]
    \centering\includegraphics[width=\textwidth]{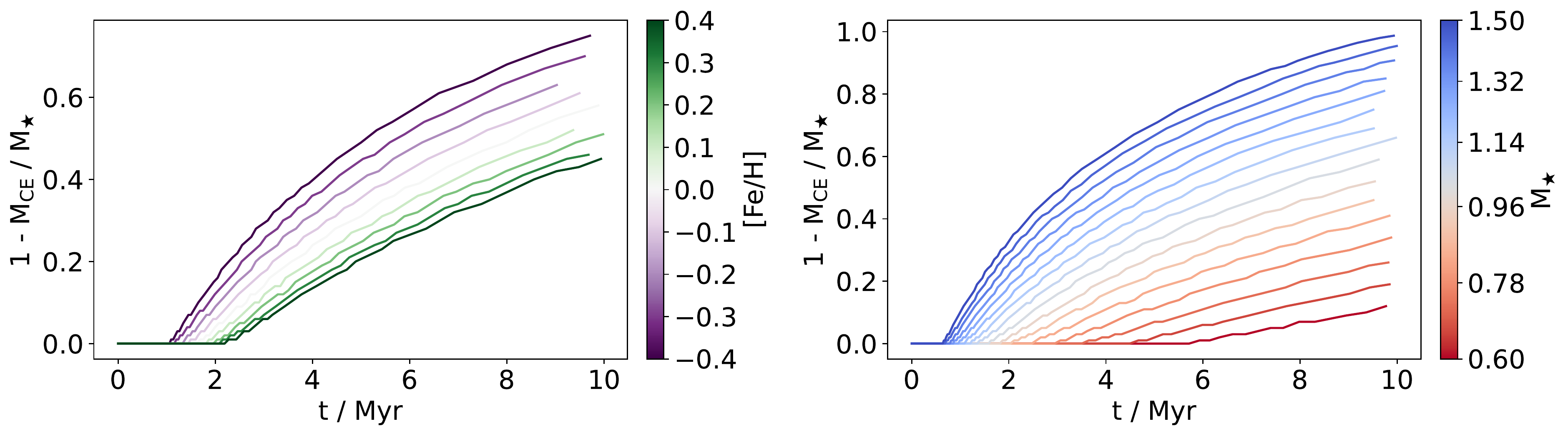}
    \caption{Time evolution of the stellar convective envelope mass obtained from models by \citet{hoppe2020} for various [Fe/H] (left panel) and $M_\star$ (right panel), indicated by a change in line color. The ordinate shows $1-M_\mathrm{CE}/M_\star$, so that if the star has a more massive convective zone, a lower value is shown, to indicate that the convective zone abundances adapt to the composition of the accretion flux more slowly.}
    \label{fig:conv_zone_time}
\end{figure*}%
For the mass evolution of the convective envelope, models from \citet{hoppe2020} are used. In essence, the mass fraction of the envelope decreases over time, allowing late accretion (or absence thereof) to have a larger impact on [X/H] than earlier on. The particular time evolution depends on two main parameters considered in this work, [Fe/H]$_\star$ and $M_\star$, which are shown in Fig. \ref{fig:conv_zone_time}. It is illustrated how for constant $M_\star$, higher stellar metallicity results in the envelope mass decreasing slower and at later times, such that the mass is higher at all times. For constant metallicity, a higher stellar mass leads to the opposite effect, where the decline of mass in the convective zone is faster and occurs earlier, so that the mass in the convective zone is lower at all times. Note that the accretion of material has an impact on the stellar evolution itself (e.g., \citealt{serenelli2011}), also modifying the convective zone evolution \citep{baraffe2010}, which is neglected for this study. Furthermore, the stellar models from \citet{hoppe2020} are used to compute the average luminosity in the first \SI{10}{\mega\year}, which in turn is used to calculate the temperature profile of the disk.

\section{Results}\label{sec:results}
\begin{table}[htp]
    \centering
    \caption[Model parameters varied in the simulations for the parameter study]{Model parameters varied in the simulations for the parameter study, and their default value.}
    \vspace*{1em}
    \begin{tabular}{lll}
    \hline
    \hline
    Parameter & Values & Default \\
    \hline
    $\alpha$ & $\num{e-4},\num{e-3}$ & \num{e-4} \\
    $u_\mathrm{frag}$ & $\SI{1}{\meter\per\second},\SI{10}{\meter\per\second}$ & \SI{1}{\meter\per\second} \\
    $M_\star$ & $\SI{.6}{\solarmass},\SI{1}{\solarmass},\SI{1.5}{\solarmass}$ & \SI{1}{\solarmass} \\\relax
    [Fe/H] & $\num{-.4},\num{0},\num{.4}$ & \num{0} \\
    \hline
    \end{tabular}
    \label{tab:parameter_space}
\end{table}%
In this section, we present simulations of the evolution of the protoplanetary disk's dust and gas components, focusing on the impact of disk accretion on the chemical abundances of the host star's convective envelope. We explore the effect of the variation of parameters described in table \ref{tab:parameter_space}. If not otherwise specified, the default values indicated in the table are used, and planetesimal formation is not considered. Furthermore, we assume a disk radial surface density profile with an exponential cutoff at $R_0=\SI{75}{\astronomicalunit}$ and a total disk mass of $M_0=M_\mathrm{disk}/M_\star=0.1$ unless specified otherwise, which results in a gravitationally stable disk.

\subsection{Abundance changes due to dust drift}\label{sec:res_nopl}
\begin{figure*}[htp]
    \centering\includegraphics[width=\textwidth]{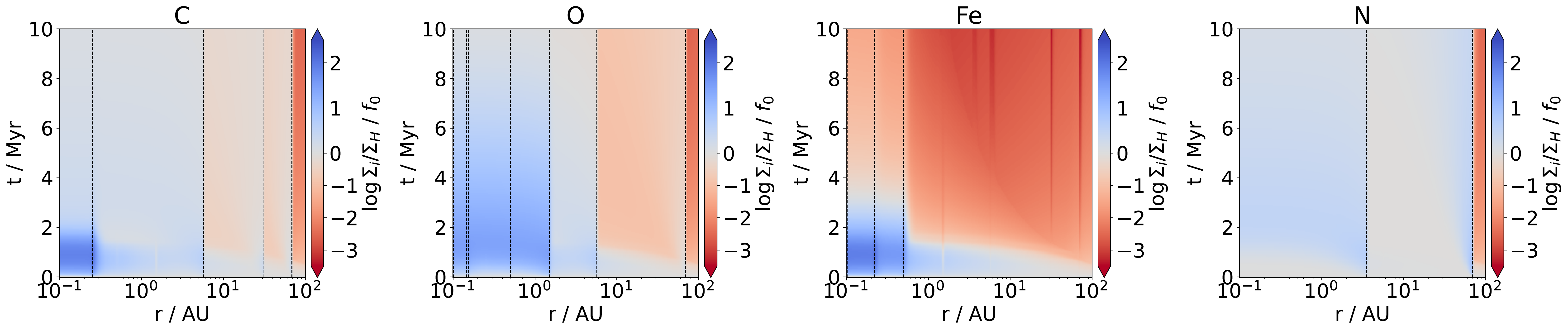}
    \caption{Surface density evolution of carbon, oxygen, iron and nitrogen, for $\alpha=\num{1e-4}$. The sum of the gas and dust component is shown. The color indicates the surface density of the element with respect to the hydrogen surface density, compared to the initial fraction at a given location on a logarithmic scale, such that a blue color indicates that the surface density of the element is enriched, and vice-versa for a red color. Due to this choice of norm, a blue color at the inner edge corresponds to an increase in [X/H] for that element up to the magnitude of the enrichment. All evaporation front locations of species where the corresponding element is a component in are indicated with black dashed lines in the relevant panels.}
    \label{fig:sigma_evo_large}
\end{figure*}%
Based on the growth and radial drift of material in the disk, the surface density of the various disk components naturally changes over time. This surface density evolution can be seen in Fig. \ref{fig:sigma_evo_large}, for $\alpha=\num{1e-4}$, where the dust grains can grow to large sizes and thus drift inwards very fast. The chemical model that we use contains numerous elements whose evolution is very similar. Therefore, the figure presents the evolution of the surface density of carbon, oxygen and nitrogen, while iron is chosen to represent all refractory species. The ice lines of all molecular species an element is part in are shown as well. The surface densities are presented relative to the surface density of hydrogen and to the initial fraction of the elements with respect to hydrogen. This achieves a view that is consistent with the analysis of [X/H] in the star's convective zone in the following sections.

As dust grains grow to large sizes, their radial drift speed surpasses the gas radial speed significantly. This leads to a pile-up of material at the ice lines, interior to which a particular species moves together with the slow gas, because it is now in a gaseous form after its evaporation. Large pebbles drift from the outer disk to the location of the ice lines quickly, and therefore create a large enrichment at those locations. One can observe that for elements mainly represented by volatile species (C, N, O), the region outside the ice line, and consequently the mass that can drift toward the ice line, is much smaller than for those present in refractory species (e.g., Fe). Altogether, there is a period of significant enrichment for refractory species close to the inner edge, which is very brief, after which the species is mostly depleted throughout the disk. In contrast, the enrichment occurs farther away from the inner edge for volatile species. Hence, the material needs to be viscously accreted in gaseous form to reach the inner edge and be removed from the disk, which means it is present also at later times in the disk. As oxygen and carbon are contained in both volatile and refractory species, they are an in-between case. This gives rise to both an early, albeit weaker, enrichment near the inner edge and a presence in the disk at later times.

\begin{figure*}[htp]
    \centering\includegraphics[width=\textwidth]{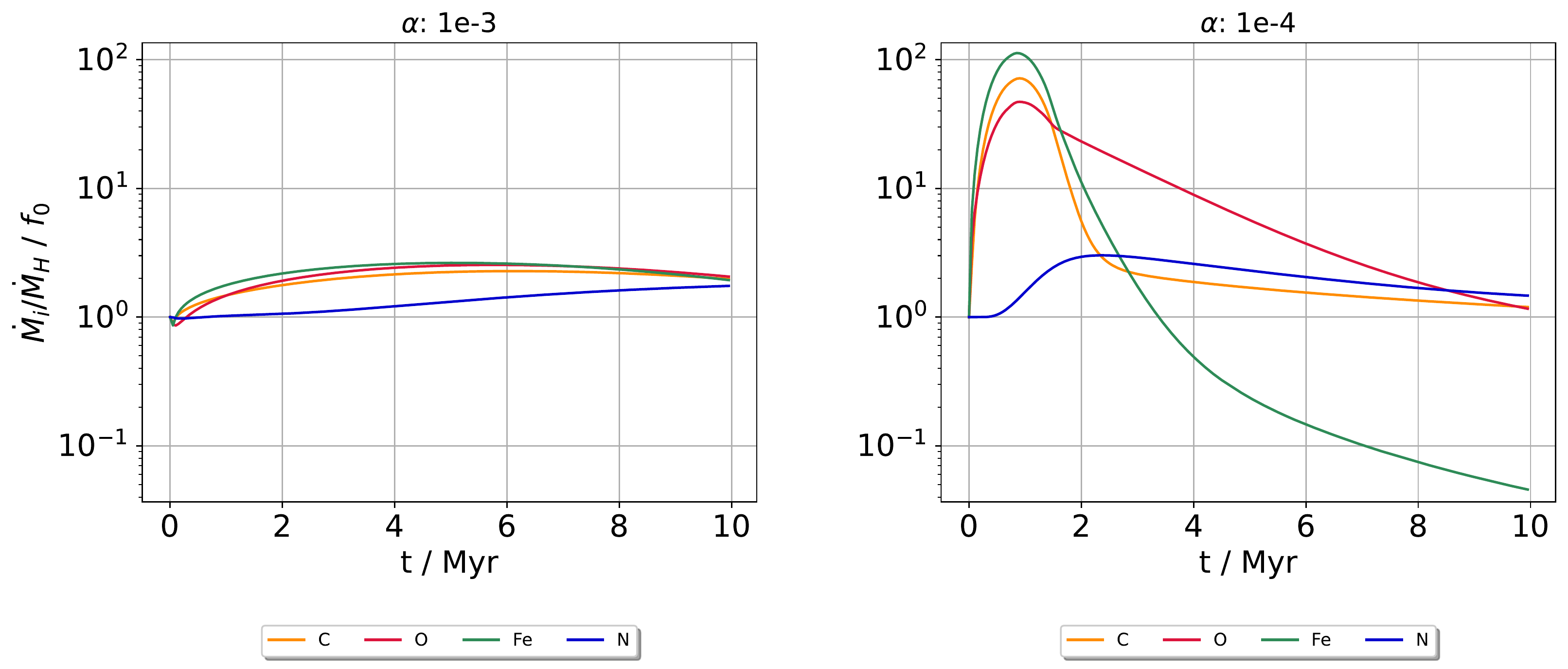}
    \caption{Accretion flux of disk material onto the host star for small particles ($\alpha=\num{1e-3}$, left panel) and for large particles ($\alpha=\num{1e-4}$, right panel) as a function of time. On the ordinate, the accretion rate with respect to the hydrogen accretion rate, $\dot{M}_i/M_H$, for various elements $i$ is shown, indicated by different colors. The accretion rate is shown normed to the mass fraction with respect to hydrogen of the element as defined by the initial condition. Therefore, if the ordinate value is larger than 1, the material that is accreted results in a growth of the abundance of element $i$ in the convective zone up to the magnitude of the enrichment. The accretion rate is directly related to the surface density evolution (see Fig. \ref{fig:sigma_evo_large}).}
    \label{fig:mdot}
\end{figure*}%
Material that crosses the inner edge is assumed to be accreted onto the star's convective zone. The difference in the surface density evolution between the elements, and its dependence on particle size, is therefore reflected in the composition of the accreted material over time, $\dot{M}$ (see Eq. \ref{eq:mdot}). This is displayed in Fig. \ref{fig:mdot}. For $\alpha=\num{1e-3}$, the slow but constant enrichment of every element but nitrogen corresponds to a similar behavior of the accretion rate, which remains largely constant, close to the initial fraction of material. For $\alpha=\num{1e-4}$, though, the enrichment of refractories corresponds to a strongly increased accretion rate at early times and a strongly decreased accretion rate at later times, where the refractories are depleted. In the case of nitrogen, the accretion rate does not increase as strongly and remains almost constant after an initial jump, owning to the slow depletion of the mostly gaseous nitrogen in a disk with low viscosity. The initial jump stems from the time needed for piled-up material at the far-out ice lines of N$_2$ and NH$_3$ to diffuse to the inner edge after evaporating.

\begin{figure*}[htp]
    \centering\includegraphics[width=\textwidth]{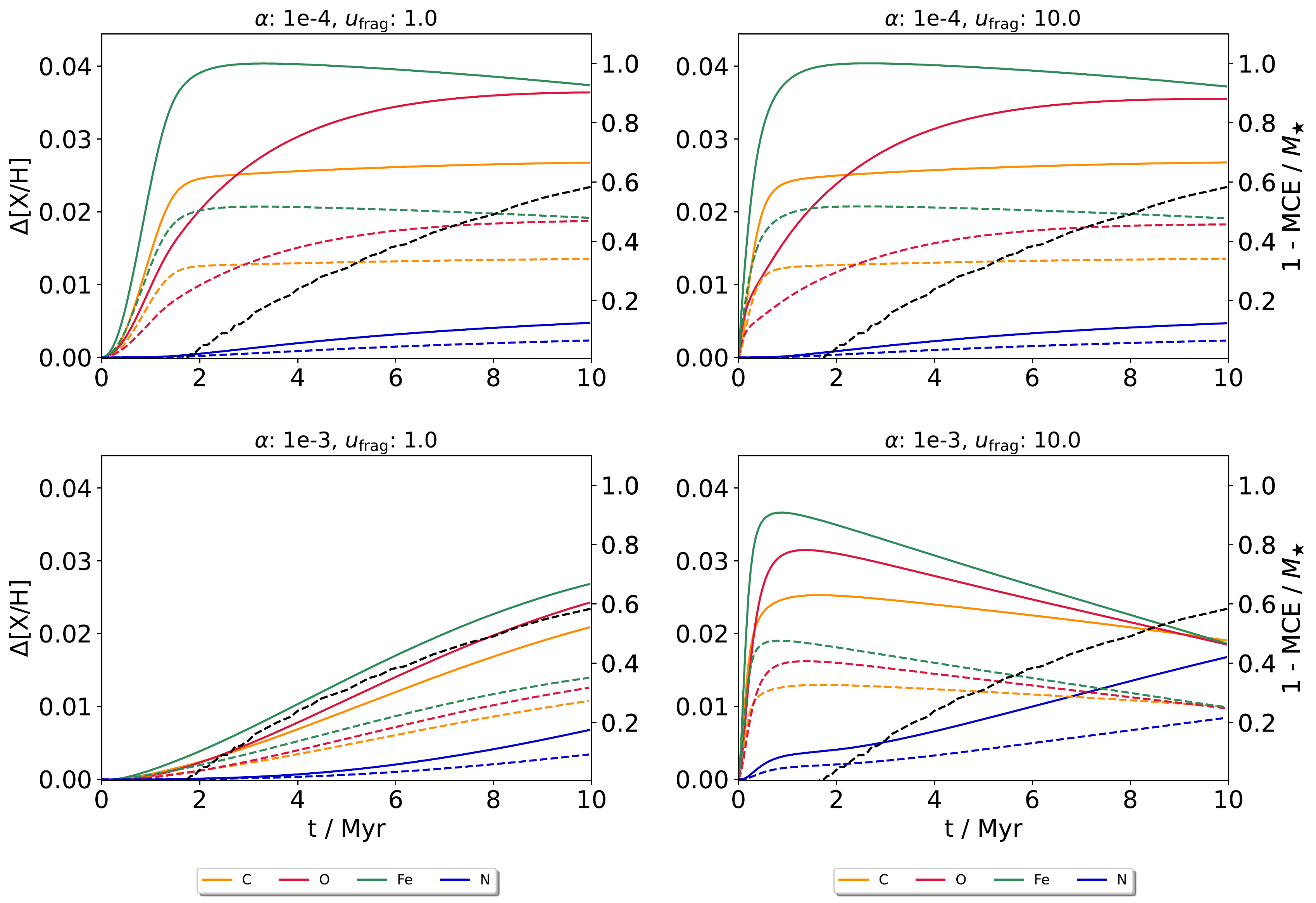}
    \caption{Stellar convective envelope elemental abundances as a function of time relative to the initial abundances, $\mathrm{[X/H]}-\mathrm{[X/H]}_0$. Four different simulations are shown, varying $\alpha$ and $u_\mathrm{frag}$ to consider different maximum grain sizes in combination with different gas accretion speeds. Differently colored solid lines represent the abundances of the elements carbon (yellow), oxygen (red), iron (green), and nitrogen (blue), while the black dashed line shows the convective zone mass evolution as $1-\frac{M_\mathrm{CE}}{M_\star}$, like in Fig. \ref{fig:conv_zone_time}, corresponding to $M_\star=\SI{1}{\solarmass}$ and $\mathrm{[Fe/H]}=0$. Solid lines indicate results for $M_0=0.1$, while dashed lines show a comparison to simulations with $M_0=0.05$.}
    \label{fig:nopla_size}
\end{figure*}%
To start, we will discuss the time evolution of the convective zone abundances caused by the composition of the accreted material while focusing on the impact of particle size and disk viscosity. Figure \ref{fig:nopla_size} shows the time evolution of $\mathrm{[X/H]}-\mathrm{[X/H]}_0$ for all four combinations of $\alpha$ and $u_\mathrm{frag}$. In addition, the evolution of the mass of the convective envelope is shown. Generally, the evolution of the chemical abundances can be explained by the envelope's [X/H] steadily approaching the value of the material accreted from the disk, with the speed of the approach being set by the mass of the envelope. For $\alpha=\num{1e-4}$ and both values of $u_\mathrm{frag}$, refractory abundance initially rises sharply, as expected in a disk that allows dust to grow to larger sizes.

Carbon abundance also rises steeply at the beginning, but falls short of reaching the same maximum abundance that the refractories and sulfur do. While the carbon grains also evaporate close to the star and are accreted quickly, only 60\% of the mass of carbon is in the form of grains, with the rest in volatile molecules accreted with the gas phase. The oxygen abundance rises less steeply than that of carbon or the refractories, coming close to the refractory abundance at the end of the simulation. The difference to carbon stems from the fact that less oxygen mass is distributed toward refractory species, which causes the less steep rise. Furthermore, the ice line of the most massive volatile component, water, is closer to the inner edge of the disk than the ice lines of the carbon-containing volatiles are. With water being present in the disk for a more extended period of time than the refractories, the lighter convective envelope at later times allows the abundance to reach a level close to the refractories toward the end of the simulation.

The difference between $u_\mathrm{frag}=\SI{1}{\meter\per\second}$ and $u_\mathrm{frag}=\SI{10}{\meter\per\second}$ in this case is the steepness of the initial slope of the abundance increase, which is greater for $u_\mathrm{frag}=\SI{10}{\meter\per\second}$; an effect that can be attributed to the difference in maximum grain size and the corresponding maximum radial drift speed. For the same reason, the maximal refractory abundance is also reached earlier for $u_\mathrm{frag}=\SI{10}{\meter\per\second}$. The time axis of Fig. \ref{fig:nopla_size} can also be viewed as the disk lifetime, so that the abscissa would indicate the final abundance difference of the central star after having accreted a disk with a lifetime as indicated on the ordinate. In turn, this interpretation leads to the conclusion that a disk that is short-lived leaves behind a star whose convective zone is more enriched in refractories and less enriched in volatiles than a long-lived disk would.

Like discussed above, the case of $\alpha=\num{1e-3},u_\mathrm{frag}=\SI{1}{\meter\per\second}$ exhibits small particles drifting slowly through the disk. Therefore, the behavior of $\Delta$[X/H] is substantially different for this case, with it increasing monotonically throughout the simulated time frame. While the dust grains still drift at speeds exceeding the gas, buildup of refractory grains is hindered, preventing the quick early accretion of grains. Instead, grains are accreted steadily as they reach their ice lines and get evaporated. As fully refractory elements are evaporated closer to the star, they can be accreted more quickly, thereby allowing the abundance difference to rise more quickly as well. Equally, oxygen and carbon rise slightly slower due to being part of both refractory and volatile species, and nitrogen rises the slowest of all considered elements as a fully volatile element.

Increasing the fragmentation velocity to $u_\mathrm{frag}=\SI{10}{\meter\per\second}$, particles can grow to larger sizes, and previous considerations can be applied. When comparing this case to the one of lower viscosity, an even steeper initial slope is shown, and the peak of the abundance difference is reached earlier on. This is because the small distance between the refractory species' ice lines and the inner edge of the disk can be traversed faster by evaporated material at this viscosity. However, there is a strong difference in evolution after the initial peak of fast refractory accretion. It relates to the state of the disk after the initial, fast growth and subsequent accretion of refractory dust has concluded. At this time, it is depleted of the elements contained in those grains (see Fig. \ref{fig:sigma_evo_large}). Moreover, gas is accreted at a high rate provided by the high viscosity. Combined, this leads to gas depleted in refractory elements being accreted rapidly, reducing the overall abundance of these species in the convective envelope. The same effect is responsible for the strong rise of nitrogen at later times, too. Nitrogen is enriched in the outer parts of the disk, and fast accretion of gas at high viscosity allows the enriched parts to be accreted onto the star during the later stages. The nitrogen abundance is higher at the end of the simulation for higher $u_\mathrm{frag}$ as well, due to a stronger enrichment at the nitrogen-related ice lines.

The relevance of late accretion in the simulation is amplified by the change in mass of the convective zone. The composition of material accreted onto to the star at later times plays a bigger role at late than at earlier times, as the shrinking convective envelope amplifies the rate of the abundance change. While this is relevant for all scenarios, it is best seen in the case of $\alpha=\num{1e-3}, u_\mathrm{frag}=10$, where the abundance of nitrogen is almost the same as the other elements at the end of the simulation.

Note that the simulation time is picked as the typical upper limit for the lifetime of protoplanetary disks \citep{ribas2015}. If the disk were to be dispersed earlier than at $\SI{10}{\mega\year}$, the abundance change would stop at that time. From this perspective, Fig. \ref{fig:nopla_size} indicates the final envelope abundances as a function of disk lifetime. Furthermore, Fig. \ref{fig:nopla_size} also presents a comparison to simulations with $M_0=0.05$. Naturally, a smaller disk mass diminishes the effect that disk accretion can have on the stellar abundances. For an even lighter disk with $M_0=0.01$, the resulting abundance differences are of the order $\Delta\mathrm{[X/H]}\sim\num{e-3}$. However, in this study, we focus on the maximal influence of disk accretion on the stellar composition and use $M_0=0.1$ for the remainder of the study.

\begin{figure*}[htp]
    \centering\includegraphics[width=\textwidth]{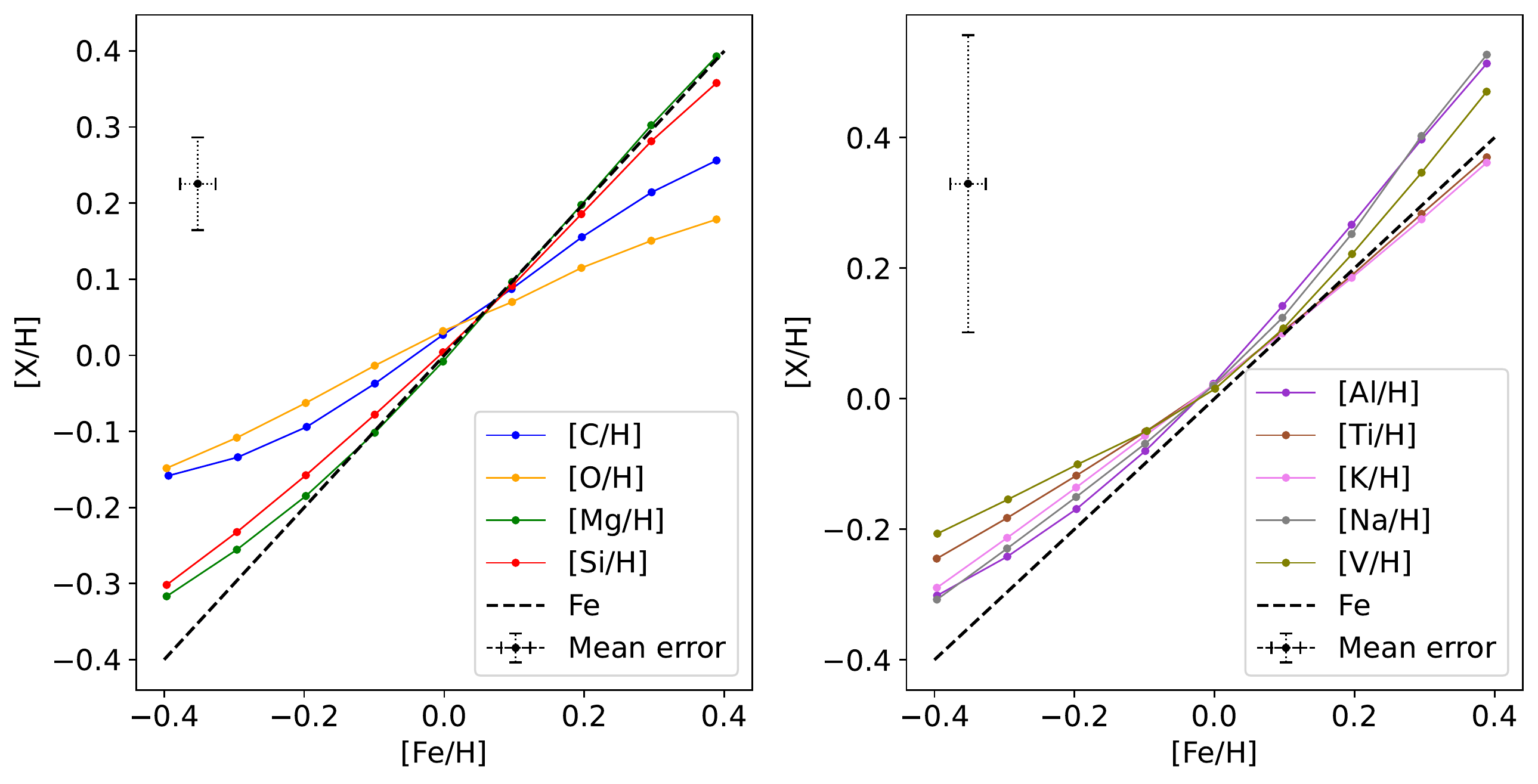}
    \caption{Scaling of initial [X/H] with [Fe/H] for the elements considered in the disk model based on GALAH+ DR3. The scaling of the abundances is indicated by solid colored lines. In addition, a reference line of 1:1 scaling is shown as a black dashed line, and a dummy data point indicating the mean error of the measurements is shown in the top left corner of each panel. In the left panel, elements that were previously considered by \citet{bitsch2020} are presented, while the remaining elements not previously considered, but used in the model, are shown in the left panel. Sulfur is not indicated; here, the same scaling as for silicon is assumed.}
    \label{fig:feh_bins}
\end{figure*}%
\subsection{Influence of different stellar metallicity and mass}\label{sec:res_nopl_stellar}
In order to discuss the influence of different initial stellar metallicities, and therefore also different initial disk elemental abundances (see Table \ref{tab:parameter_space}), it is necessary to consider how a change in metallicity affects the abundances of the chemical species. While typically the iron abundance [Fe/H] is used as a proxy for the overall metallicity, the elemental abundances scale with [Fe/H] separately for each element (e.g., \citealt{burbidge1957,bitsch2020}). Analogously to the methods employed in \citet[BB20]{bitsch2020}, the scaling of abundances of the chemical model's elements is obtained from the GALAH+ survey's third data release \citep{galah2021}. Due to changes in the reduction pipeline and analysis workflow between the second and third data release, the selection criteria were adjusted compared to BB20. We require a star to have $T_\mathrm{eff}>\SI{5000}{\kelvin}$ and $\log g > 4$, as well as quality flags of $\texttt{flag\_sp}=0$, $\texttt{flag\_fe\_h}=0$, $\texttt{flag\_x\_fe}=0$ and $\texttt{snr\_c3\_iraf}> 30$. Like in BB20, stars are grouped in [Fe/H] bins with size 0.1. The corresponding abundance [X/H] is then obtained as the mean [X/H] per bin, shown in Fig. \ref{fig:feh_bins}. For sulfur, the same scaling as for silicon is assumed \citep{chen2002}. While the uncertainty of the elemental abundances is large, especially for elements not considered by BB20, the resulting scaling of the elements is employed whenever [Fe/H] is adjusted in a simulation. We note that there are slight differences in the error bars for the individual elements between our analysis using GALAH DR3 compared to BB20, who used GALAH DR2. In addition, the error bars and trends are also recovered with the Apogee survey, see \citet{cabral2023} for a comparison.

\begin{figure*}[htp]
    \centering\includegraphics[width=\textwidth]{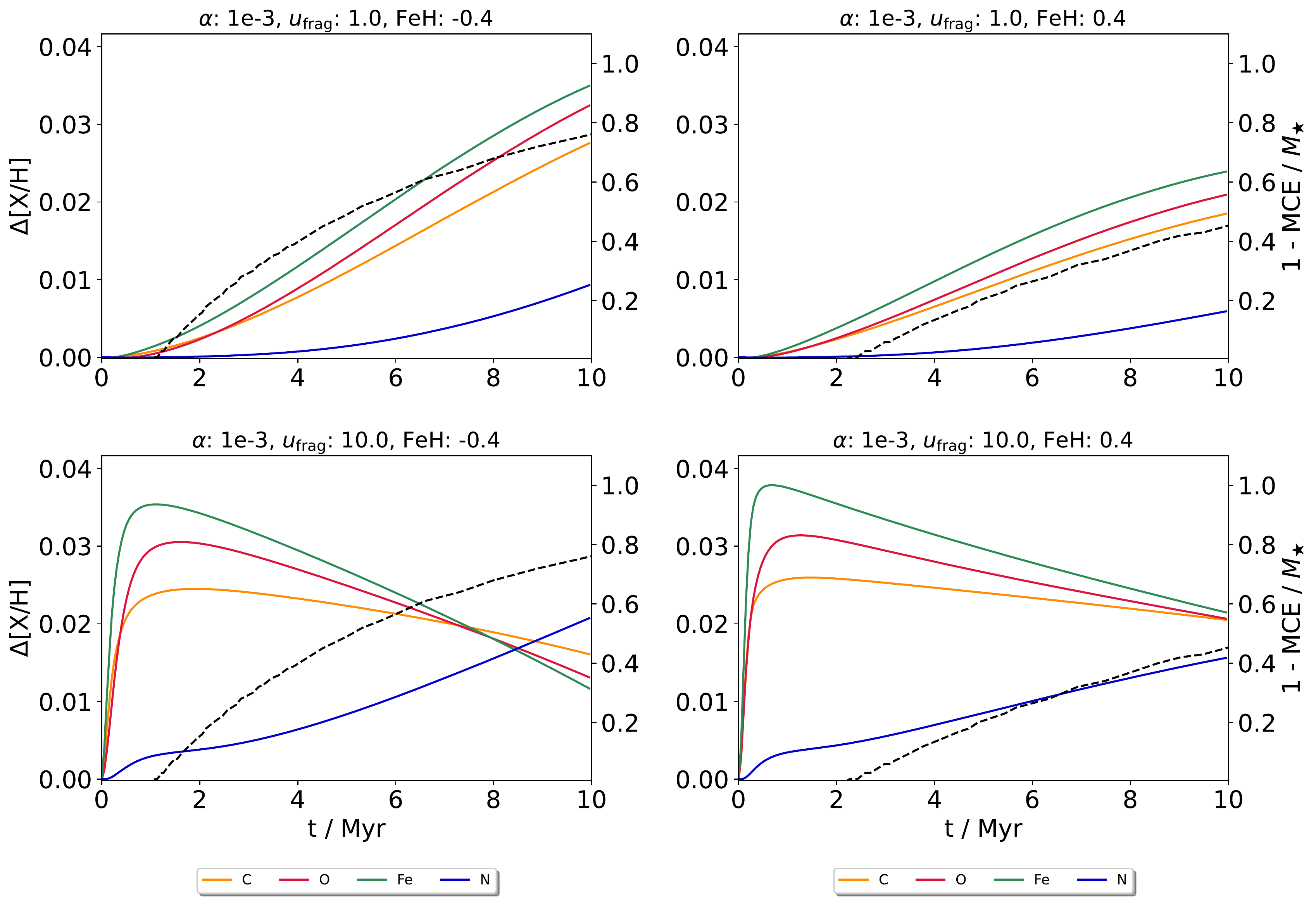}
    \caption{Like Fig. \ref{fig:nopla_size}, but the initial [Fe/H] is varied instead of $\alpha$. It is kept constant at $\alpha=\num{1e-3}$. On the left-hand side, the case for an initial metallicity of $\mathrm{[Fe/H]}=-0.4$ is shown for both $u_\mathrm{frag}=\SI{1}{\meter\per\second}$ and $u_\mathrm{frag}=\SI{10}{\meter\per\second}$, whereas the right-hand size shows simulations with $\mathrm{[Fe/H]}=+0.4$.}
    \label{fig:nopla_feh}
\end{figure*}%
The time evolution of the convective zone abundances for low metallicity ($\mathrm{[Fe/H]}=-0.4$) and high metallicity ($\mathrm{[Fe/H]}=0.4$) are shown in Fig. \ref{fig:nopla_feh} for two values of $u_\mathrm{frag}$. When considering the results of the simulations with $u_\mathrm{frag}=\SI{1}{\meter\per\second}$, where grains cannot grow to large sizes, the final values for [X/H] are smaller for higher initial metallicity, while the overall trend remains similar. In contrast, simulations with large dust grains show larger final abundances for a higher metallicity. In addition, the species that exhibits the strongest enrichment at the end of the simulation changes between refractories having the highest abundance for $\mathrm{[Fe/H]}=+0.4$ and nitrogen, carbon and oxygen having the highest abundance for $\mathrm{[Fe/H]}=-0.4$.

Since the initial metallicity is changed for both the disk and the star, both being made out of material from the same molecular cloud, there are several changes to the disk evolution when compared to solar metallicity. First, the total surface density of dust in the disk increases for a higher metallicity, which decreases the growth timescale of dust grains. Second, a higher metallicity star has a lower luminosity averaged over \SI{10}{\mega\year}, as computed from the \citet{hoppe2020} stellar models. As the outer region's temperature profile is dominated by stellar irradiation, this leads to a colder outer disk, shifting ice lines of volatile molecular species closer to the star. At the same time, the inner disk becomes hotter, as the dust optical depth increases with the dust-to-gas ratio. Third, as the elements scale differently with [Fe/H] (see Fig. \ref{fig:feh_bins}) and the relative distribution of elements determines the partitioning in the chemical model as per table \ref{tab:partitioning}, mass contributions to elements by chemical species can shift to more volatile or refractory species, for example, decreasing the water fraction by a factor of ${\sim}2$ when increasing [Fe/H] from $-0.4$ to 0.4.

However, the difference in the evolution of the convective zone abundances is governed by the altered convective zone evolution. Material accreted late during the disk's lifetime has a particularly strong impact on the changes caused by the metallicity variation. Figure \ref{fig:conv_zone_time} shows that the convective zone mass fraction is larger for all times when increasing the metallicity. Therefore, the rate of abundance change is lowered during late accretion phases for high metallicities. As disks that do not allow significant growth of dust grains do not exhibit a drop in abundance due to the accretion of depleted material, this results in smaller final abundances in such a case. If grains can grow larger and depletion becomes relevant, the drop of abundances caused by the accretion of depleted material is diminished by the more massive convective envelope. In such cases, high metallicities create large final abundances.

\begin{figure*}[htp]
    \centering\includegraphics[width=\textwidth]{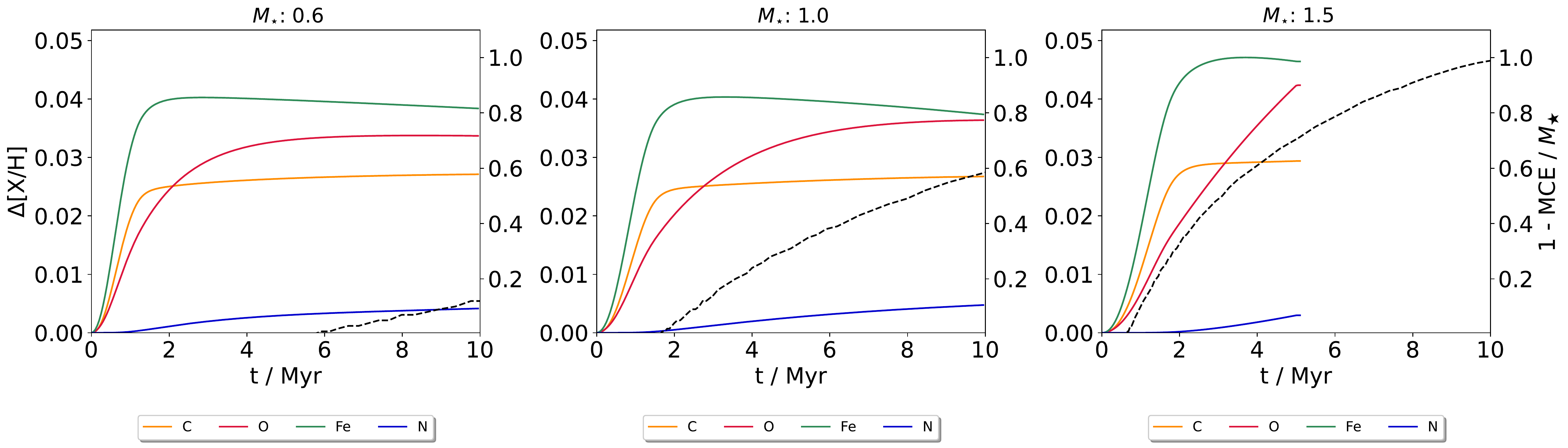}
    \caption{Like Fig. \ref{fig:nopla_size}, but the stellar mass is varied instead of $\alpha$ and $u_\mathrm{frag}$. The three panels show simulations with $M_\star\in\{\SI{0.6}{\solarmass},\SI{1.0}{\solarmass},\SI{1.5}{\solarmass}\}$. For the simulation with $M_\star=\SI{1.5}{\solarmass}$, the simulation run time is reduced to \SI{5}{\mega\year} in accordance with observations of accretion disks around massive stars.}
    \label{fig:nopla_mstar}
\end{figure*}%
If the central star is more massive, the mean luminosity is higher over the disk lifetime, while the convective envelope mass is reduced (see Fig. \ref{fig:conv_zone_time}). Both of these aspects are also modified by a change in initial [Fe/H], making the discussion of the $M_\star$ parameter analogous to the discussion of metallicity changes. With the disk mass being defined relative to the stellar mass (see Table \ref{tab:parameter_space}), the surface density increases for increasing stellar mass, while keeping the same dust-to-gas ratio, unlike for a change in metallicity. Similar to the case of metallicity variation, changes to the abundance evolution are dominated by the change in convective zone size.

Figure \ref{fig:nopla_mstar} shows the abundance changes for stellar masses of \SI{0.6}{\solarmass}, \SI{1}{\solarmass} and \SI{1.5}{\solarmass}. For the highest stellar mass, the disk lifetime was shortened to \SI{5}{\mega\year} to account for observations of protoplanetary disks around more massive stars \citep{ribas2015}, as well as to allow the star to still have a convective envelope at the end of the disk's life. For refractory elements, there is only a small difference between the cases of a low mass and a solar mass star. This is because the refractory dust grains are accreted very early, at a time when the star is still fully convective in both cases. On the other hand, for the high mass star, the early accretion of refractories is still commencing as the mass of the convective zone shrinks, allowing for a higher abundance peak. For the volatile nitrogen and the large water component of oxygen, the accretion is prolonged over the disk's lifetime. Therefore, the presence of a more massive convective zone at later times prevents those elements from achieving an abundance difference as high as for the solar mass star. As the most massive star has the smallest envelope, the opposite is true here, where oxygen can reach higher values of $\Delta\mathrm{[X/H]}$.

\subsection{Influence of planetesimal formation}\label{sec:res_nopl_planetesimal}
\begin{figure*}[htp]
    \centering\includegraphics[width=\textwidth]{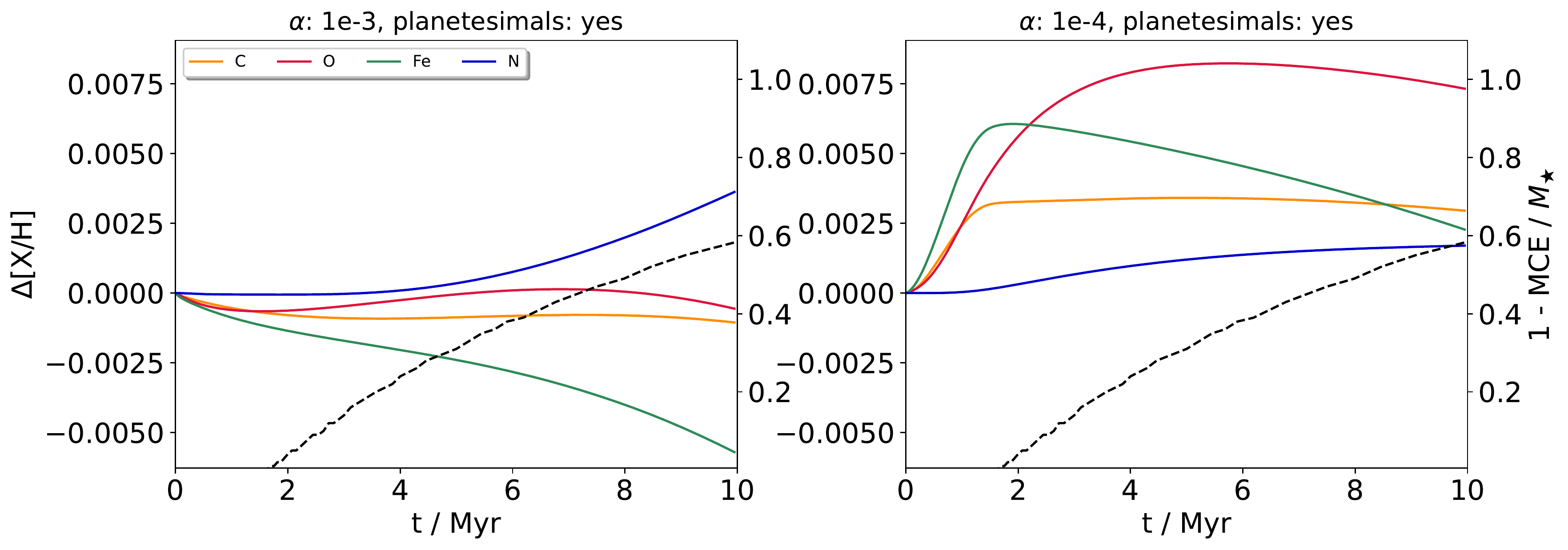}
    \caption{Like Fig. \ref{fig:nopla_size}, but only $\alpha$ is varied. For the shown simulations, planetesimals are formed in the disk.}
    \label{fig:nopla_plas}
\end{figure*}%
We will conclude the discussion of planet-less accretion by considering the formation of planetesimals in the disk. Since planetesimals form from a fraction of the pebble flux, dust locked up in planetesimals cannot be accreted onto the star, as planetesimals do not experience radial drift. Therefore, the final stellar abundances are lower if planetesimal formation is included. Two examples, for high and low viscosity, are shown in Fig. \ref{fig:nopla_plas}. With the direct dependence of the planetesimal formation sink term on the pebble flux, the absolute speed of the dust grains sets how strongly the enrichment is diminished, rather than the speed difference to the gas that is responsible for the enrichment itself. Hence, at high viscosity $\alpha=\num{1e-3}$, even though the dust grains do not grow to large sizes, a large mass of heavy elements gets locked up and prevented from contributing to enrichment. As, in this case, the over-densities at the ice lines are already small, this results in a negative value of $\Delta\mathrm{[X/H]}$ at the end of the simulation for refractory species. In disks with lower viscosity, however, the convective envelope is still enriched in refractories after \SI{10}{\mega\year}, as strong enrichment is still possible even if some material is locked up. For all simulations presented in Fig. \ref{fig:nopla_plas}, the inclusion of planetesimal formation results in a maximal abundance difference for all elements below \num{e-2}, an order of magnitude lower than differences obtained without it, which is lower than the precision that can be achieved with observations at this time.

Planetesimals can be formed anywhere in this model, and the total pebble flux is the largest for refractory species because they are in solid form even in the inner part of the disk. Therefore, they are subject to more removal of material. This is reflected in iron being the most depleted at the end of the simulation. It is also apparent when comparing the final abundances of oxygen and carbon. In our model, 60\% of carbon is in the form of grains, whereas most oxygen is in water (see Table \ref{tab:partitioning}). Since the carbon grain evaporation front is closer to the star than the water ice line, more carbon is locked up in planetesimals than oxygen. Increasing the planetesimal formation efficiency $\epsilon$ would shift the location of planetesimal formation further outward, as pebbles in the outer disk would be converted into planetesimals before they can contribute to formation in regions of the disk closer to the star \citep{voelkel2020}. Consequently, this means that planetesimals would include an even higher relative mass fraction of solids with lower condensation temperatures, as opposed to the less efficient case where the formation would be shifted more toward the inner disk regions, where they are present in gaseous form.

\section{Influence of a giant planet}\label{sec:res_pl}
In this section, simulations that include the formation of one giant planet, accreting pebbles onto a protoplanetary seed to from its core and accreting gas after reaching pebble isolation mass, if applicable, are presented. A brief summary about the employed planet formation model is given in Appendix \ref{sec:meth_peb_iso}.

\subsection{Planet location}\label{sec:res_pl_loc}
The placement of the planet will be varied to study how blocking different molecular species in the disk from accreting affects the final abundances. To achieve consistency with disks with different temperature profiles as model parameters are varied, the planet is always placed just outside one of three ice lines, i.e. at an orbital radius 110\% of the ice line's radial distance to the star. The considered ice lines are those of Fe$_3$O$_4$, to study a case where only highly refractory elements are blocked, H$_2$O and CO$_2$. For simplicity, the planet is prevented from migrating.

\begin{figure*}[htp]
    \centering\includegraphics[width=\textwidth]{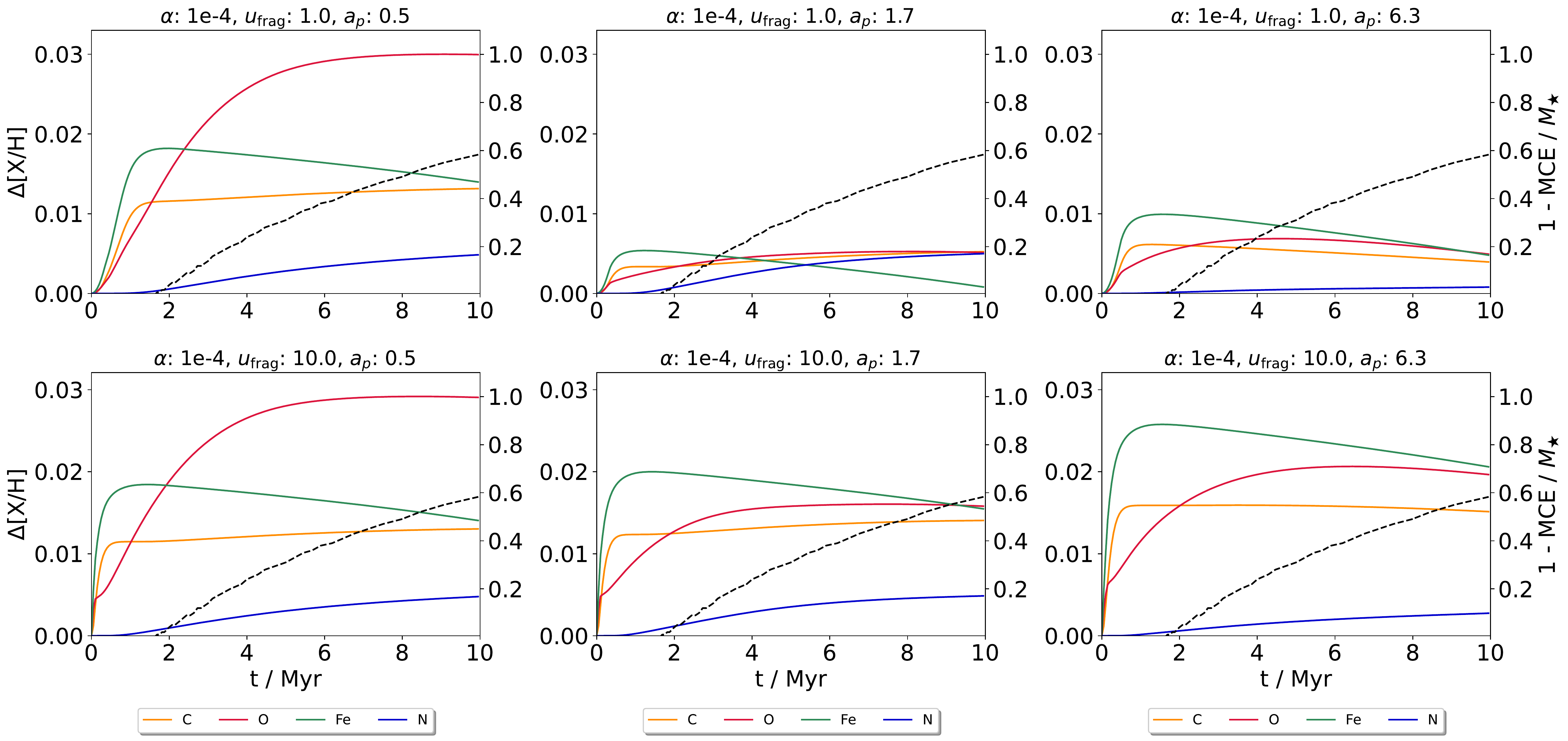}
    \caption{Like Fig. \ref{fig:nopla_size}, but now, a planet is included in the disk. The viscosity is kept constant, but $u_\mathrm{frag}$ is varied. Simulations where the planet was placed at three different locations are shown, relative to the Fe$_3$O$_4$, H$_2$O and CO$_2$ ice lines. Each column represents a different planet placement, with it being placed further out going from the left-most to the right-most column. The top row depicts disks with $u_\mathrm{frag}=\SI{1}{\meter\per\second}$ and the bottom row those with $u_\mathrm{frag}=\SI{10}{\meter\per\second}$.}
    \label{fig:pla_xh_ap}
\end{figure*}%
\begin{figure*}[htp]
    \centering\includegraphics[width=\textwidth]{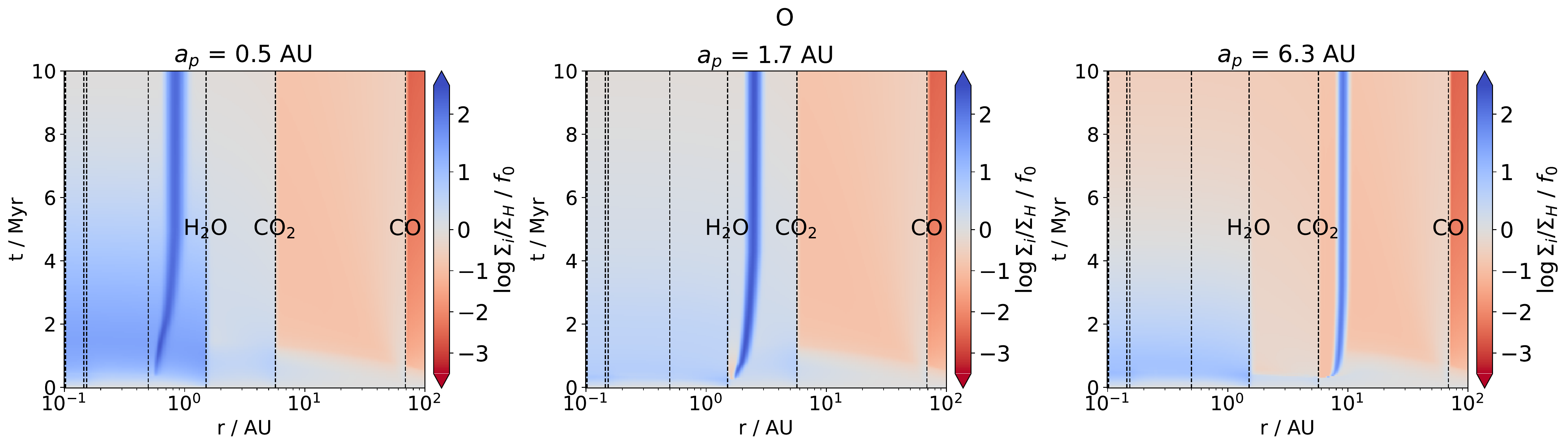}
    \caption{Like Fig. \ref{fig:sigma_evo_large}, but the surface density evolution of oxygen in solid and gaseous form is shown for three different planet formation locations. Additionally, the H$_2$O, CO$_2$ and CO ice lines are labeled.}
    \label{fig:planet_sigma_evo}
\end{figure*}%
Once the giant planet reaches the pebble isolation mass, solid material weakly coupled to the gas is trapped in a pressure bump outside the planet's orbit and prevented from moving inward, which blocks its accretion onto the star. Hence, the position of the planet relative to the ice lines impacts the composition of material that is added to the envelope. Figure \ref{fig:planet_sigma_evo} shows the pile-up of oxygen-containing solids outside the planet's orbit. If the planet forms at the innermost location, only oxygen contained in refractory components is blocked, whereas a larger fraction of oxygen is prevented from accreting onto the star if the water and CO$_2$ ice lines are blocked.

The impact of the planet's positioning is reflected in the time evolution of [X/H] of the convective zone, shown in Fig. \ref{fig:pla_xh_ap}. The figure displays both the three different planet positions relative to the aforementioned ice lines, and cases for both $u_\mathrm{frag}=\SI{1}{\meter\per\second}$ and $u_\mathrm{frag}=\SI{10}{\meter\per\second}$, as a proxy for the dust grain size. Most notably, the position of the planet relative to the water ice line is reflected in the final oxygen abundance in the convective zone. For the innermost planet, this abundance is highest, as the water ice line is exterior to the planet's orbit. For the other two farther outward planets, the opposite holds true, reducing the final abundance.

While the planet's location relative to the ice lines defines what chemical species can potentially be blocked, some material can still be accreted. Material present interior to the planet before core formation is always accreted onto the star. If the core forms at a large distance to the star, more material is unaffected and can contribute to the abundance increase in the convective envelope. Furthermore, the growth time of the planet is prolonged in the inner regions of the disk due to a reduced pebble flux. Also, the magnitude of the isolation mass increases with radial distance \citep{lambrechts2014,bitsch2018a}. As a result, a planet that forms furthest from the star takes longest to reach pebble isolation, and more grains can be accreted before that happens, creating a larger abundance in the envelope. The middle planet takes the least time to reach the isolation mass, whereas the innermost planet's time lies in between the other two cases. In disks where grains remain small, this difference is most noticeable. Overall, the magnitude of the pebble flux is indicative of both the growth time of the planet and the total grain mass that drifts past the planet before reaching the isolation mass.

Disks with small grains form less massive planets. The pressure bump that is created by the planet to reach pebble isolation is therefore weaker. If the planet is placed at the innermost location, the low pebble isolation mass gives rise to a particularly weak pressure bump. Combined with significant gas drag experienced by small particles, this low pressure bump does not successfully prevent the grains from drifting past the planet's orbit, even after the pebble isolation mass is reached. This is reflected in Fig. \ref{fig:pla_xh_ap}. Comparing the top left and top center panels, the former shows a significantly stronger enrichment of, for example, iron. All iron related ice lines are blocked in both cases. Here, the stronger enrichment is caused by the failure of the weak pressure bump to prevent pebble drift past its orbit.

\subsection{Implications for Jupiter and the Solar System}\label{sec:res_jup}
Another interesting application of the here presented methods is to study how Jupiter's growth and the accretion of the Solar protoplanetary disk influenced the composition of the Sun. We do not aim for great detail in this section, but to give order-of-magnitude estimates using a simple approach.

As the most massive planet of the Solar system, Jupiter has the largest impact not only on the dynamics, but also on planet-disk interactions and potential influence on the accretion of solids onto the Sun. Since the \texttt{chemcomp} code can only consider a single planet in the disk, it is the only planet considered for determining the values of $\Delta\mathrm{[X/H]}$ expected for the Sun. Several further simplifications are made. First, Jupiter is kept fixed at its present-day location of ${\sim}\SI{5.2}{\astronomicalunit}$, neglecting migration scenarios where it has experienced both significant inward and outward migration in conjunction with Saturn \citep{walsh2011}. Second, the gas accretion of the planetary core is stopped once it reaches the mass of Jupiter. Changing the accretion rate of gas by a constant factor $\geq 0.1$ was found to have no impact on the result. Furthermore, simplifications have been made for the initial conditions of the Sun and its disk. In the previous sections, it was shown that the initial disk and star abundances are different from the present-day values after the disk has been accreted in part and dispersed. Despite this, rather than making assumptions about the initial abundances of the Sun and trying to reproduce present-day Solar values, the simulation uses the observed Solar values (e.g., \citealt{asplund2009}) as a starting point to find the abundance changes created by disk accretion. The main difference to the real situation created by this approach is the starting metallicity, which sets the evolution of the convective zone mass and the initial heavy element surface density in the disk. In Sects. \ref{sec:res_nopl} and \ref{sec:res_pl}, it was shown that the differences created by disk accretion are of the order of \num{e-2} in [X/H], which can be viewed as the expected difference in metallicity for the stellar model. The metallicity resolution of the stellar models used in \citet{hoppe2020} is 0.05 dex, so that the change to the Solar model by considering a modified initial metallicity of the Sun would in most cases be outside the scope of the model resolution.

\begin{figure*}[htp]
    \centering\includegraphics[width=\textwidth]{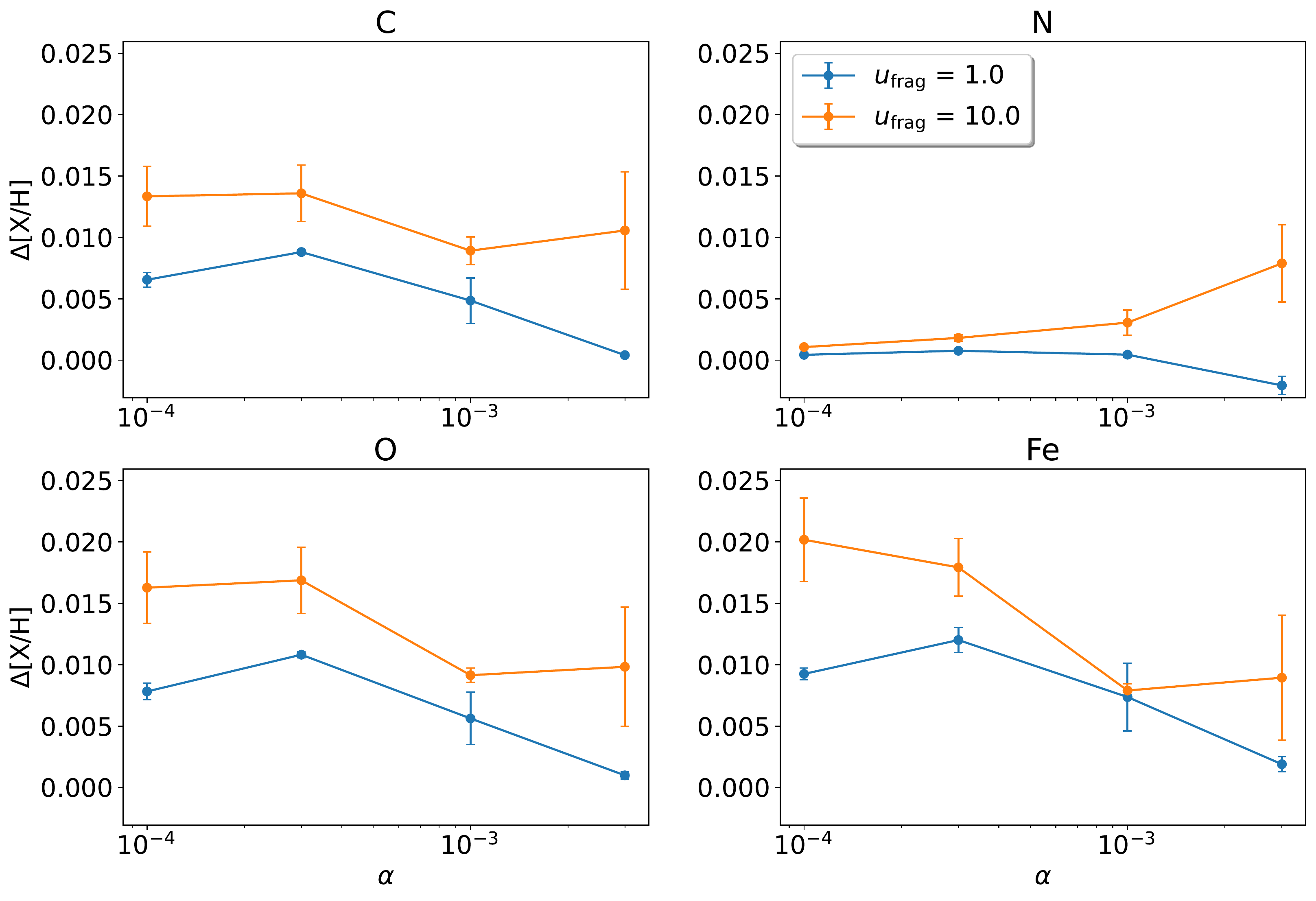}
    \caption{Solar convective zone abundance changes created by accretion of a disk forming Jupiter as a function of disk viscosity $\alpha$ for $u_\mathrm{frag}=\SI{1}{\meter\per\second}$ (blue) and $u_\mathrm{frag}=\SI{10}{\meter\per\second}$ (orange). The data points show the mean value for simulations with $R_0\in\{\SI{75}{\astronomicalunit},\SI{150}{\astronomicalunit}\}$ and $M_0\in\{0.05,0.1\}$, and the error bars indicate the standard deviation.}
    \label{fig:jup_summary}
\end{figure*}%
The expected change of the convective zone abundances is presented in Fig. \ref{fig:jup_summary}, where the final abundance difference after \SI{4.5}{\mega\year} is shown. The choice if disk lifetime is based on information from the formation of chondrites \citep{johnson2016,kleine2020}. The abscissa shows the results for four values of $\alpha$ sampling the log space between $\num{e-4}$ and $\num{3e-3}$ regularly. Again, cases for low ($u_\mathrm{frag}=\SI{1}{\meter\per\second}$) and high ($u_\mathrm{frag}=\SI{10}{\meter\per\second}$) fragmentation velocity are considered. In the applied model, Jupiter blocks the water ice pebbles, but not the CO$_2$ pebbles, because they have already evaporated upon reaching Jupiter's orbit. Note that Fig. \ref{fig:jup_summary} shows the abundance difference compared to the initial Solar value, not to a case where Jupiter does not form in the disk.

The disk cutoff radius $R_0$ and mass $M_0$ affect the evolution of the disk, mainly due to a change in the surface density and total dust mass. For higher $M_0$, the higher available mass of dust creates a stronger enrichment at the ice lines, which in turn is reflected by the convective zone abundances reaching a higher peak value. A smaller $R_0$ increases the dust surface density $\Sigma_d$. While this increases disk temperature, the dominant effect is the reduced dust growth and drift timescale. As a result, the accretion flux onto the central star becomes enriched already at earlier time, when the convective zone is still massive. Depending on the subsequent convective zone evolution, given by stellar mass and metallicity, as well as the general accretion timescale, given by $\alpha$ and $u_\mathrm{frag}$, this can increase or decrease the final and peak elemental abundances. Therefore, final Solar abundance differences caused by disk accretion are computed for $R_0\in\{\SI{75}{\astronomicalunit},\SI{150}{\astronomicalunit}\}$ and $M_0\in\{0.05,0.1\}$. The expected change of abundance is then taken as the average.

Because nitrogen-related ice lines are outside Jupiter's orbit, [N/H] is not affected by the pressure bump caused by it. As such, the nitrogen abundance in the Sun's convective envelope is governed by disk effects. At low viscosity, the pile-up at the N$_2$ ice line is not accreted onto the star. Combined with the fact that only 10\% of the total nitrogen mass is in the form of NH$_3$, changing the size of the particles without changing the viscosity, i.e. due to variation of $u_\mathrm{frag}$, has little effect. This changes once the viscosity is high enough for the enrichment at the N$_2$ ice line to reach the inner edge of the disk during the simulation. In that case, a stronger enrichment corresponding to a larger fragmentation velocity results in a higher nitrogen abundance. However, for $u_\mathrm{frag}=\SI{1}{\meter\per\second}$ and $\alpha=\num{3e-3}$, the accretion of gas is more efficient than evaporation of material at the nitrogen ice lines, where no significant pile up occurs due to small grain sizes. In this scenario, the gas becomes depleted in nitrogen and the final value for [N/H] is lower than initially.

For the other elements, though, at least some molecules that contain them have their ice line blocked once the planet carves the shallow gap in the disk. The time it takes for Jupiter to grow to pebble isolation mass and the mass of material that drifts by its orbit before that is indicative for the final elemental abundances, as discussed above. The relative change in time scales when raising the fragmentation velocity allows more material to pass the planet's orbit and accrete onto the star then for a low $u_\mathrm{frag}$ (see Fig. \ref{fig:pla_xh_ap}). In the high $u_\mathrm{frag}$ case, Jupiter always reaches the isolation mass. For high disk mass, it does so faster for higher viscosity until $\alpha=\num{1e-3}$. For $\alpha=\num{3e-3}$, the combined effect of a high pebble flux and smaller pebble sizes lead to Jupiter growing for longer before the isolation mass is reached. If the disk mass is lower, it takes Jupiter a much shorter amount of time to reach isolation for $\alpha=\num{3e-3}$, but the difference between light and massive disks decreases as the viscosity decreases. For $\alpha=\num{1e-3}$, the planetary core is able to reach the isolation mass earlier to block refractories if the disk is massive. The trends of isolation mass timing are directly reflected by a higher or lower final abundance of elements with refractory components, respectively. In addition, the iron abundance drops more significantly for cases where Jupiter is quick to form the pressure bump, because all ice lines of solids that contain those elements are inside the planets orbits, unlike for carbon or oxygen.

A value of $u_\mathrm{frag}=\SI{1}{\meter\per\second}$ does not always allow Jupiter to grow to the pebble isolation mass during the disk's lifetime. In fact, for $\alpha\geq\num{1e-3}$, regardless of disk mass or size, a partial gap to block solid grains is not opened. This is related to the fact that the gap opening is more difficult in discs with larger viscosity (e.g. \citealt{crida2006,kanagawa2018,bergezcasalou2020}). The final abundances of all elements present in refractory species for those high viscosities therefore show similar values, caused just by the growing core of Jupiter accreting part of the pebble flux. Otherwise, the final abundance is created by disk evolution effects independent of a planet (see Sect. \ref{sec:res_nopl}). With dust growth being negligible for $\alpha=\num{3e-3}$, no enrichment of material takes place in the disk and the abundance in the convective envelope remains at Solar value. For low viscosity, Jupiter reaches pebble isolation, albeit late during the disk's lifetime. At $\alpha=\num{3e-4}$, Jupiter takes long to create a pressure bump, so the abundance of refractories in the envelope is higher in those cases.
\section{Application to HD106515}\label{sec:res_hd}
Finally, the aim of this section is to compare abundance differences created by planet formation and disk evolution to observations and study whether differences between wide binaries can be recreated. In particular, the wide binary system HD106515 is considered. It consists of two solar-like stars with a semi-major axis of $345^{+95}_{-47}\ \si{\astronomicalunit}$ \citep{rica2017}. The primary constituent HD106515 A has a mass of $0.888\pm 0.018\ \si{\solarmass}$ and a metallicity of $\mathrm{[Fe/H]}=+0.016\pm 0.009$, while the companion HD106515 B has a mass of $0.861\pm 0.015\ \si{\solarmass}$ and a metallicity of $\mathrm{[Fe/H]}=+0.022\pm 0.010$ \citep{saffe2019}.

\begin{figure}[htp]
    \centering\includegraphics[width=\linewidth]{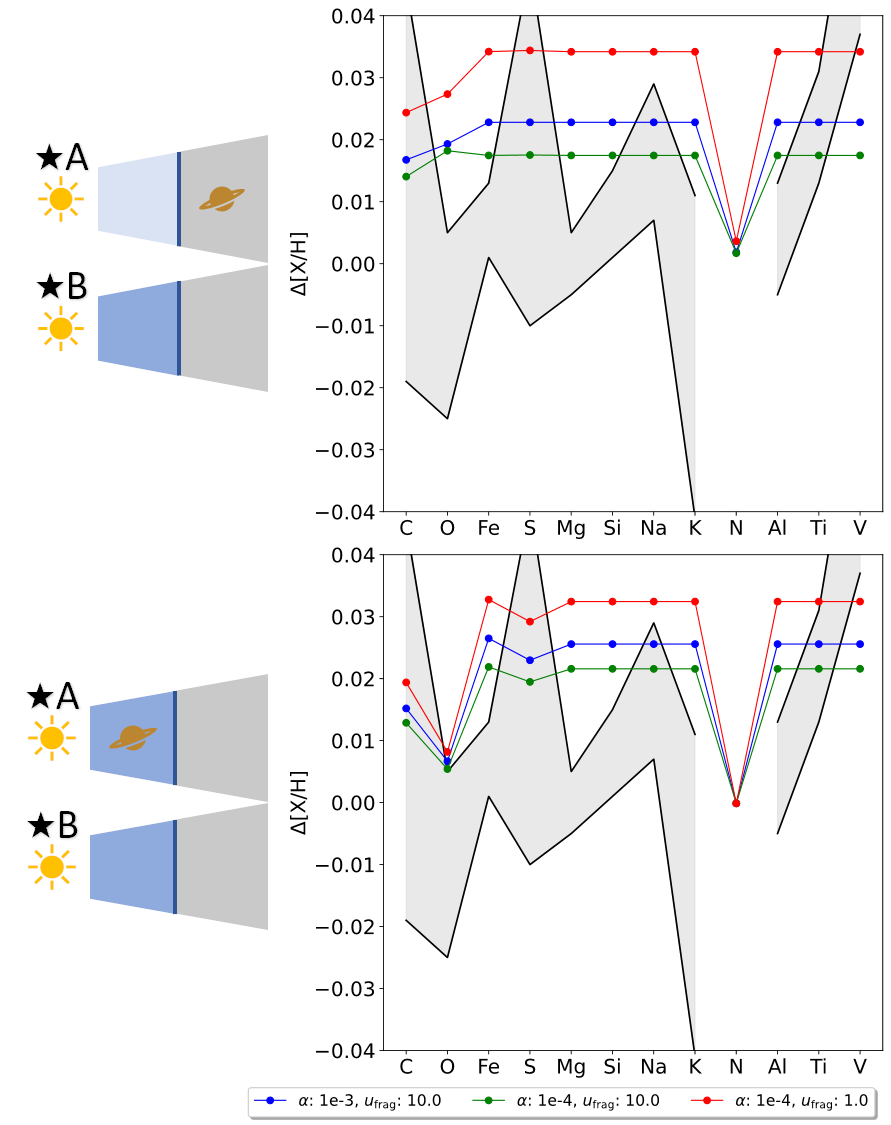}
    \caption{Abundance differences between HD106515 B and A caused by planet formation taking place only around star A. The top shows the case of the planet forming where it is observed today at \SI{4.5}{\astronomicalunit}, which lies outside the water ice line, while the bottom considers the planet forming inside the water ice line, just outside $r_{\mathrm{Fe}_3\mathrm{O}_4}$. The gray area delimited by black lines corresponds to the measurements including the $1\sigma$ error \citep{liu2021}. In blue, the disks around the constituents are parameterized by $\alpha=\num{1e-3},u_\mathrm{frag}=\SI{10}{\meter\per\second}$. The red and green lines describe disks with a viscosity of $\alpha=\num{1e-4}$ with $u_\mathrm{frag}=\SI{1}{\meter\per\second}$ and $u_\mathrm{frag}=\SI{10}{\meter\per\second}$, respectively.}
    \label{fig:hd_summary_pl}
\end{figure}

Observations of this system show that the two binary stars do not have the same elemental abundances. A clue to the origin of this difference could lie in the fact that the primary star harbors a confirmed giant planet, HD106515A b. Its mass is determined to be $18.9^{+1.5}_{-1.4}\ \si{\jupitermass}$, and it has an observed semi-major axis of $4.48\pm 0.05\ \si{\astronomicalunit}$. In Sect. \ref{sec:res_pl}, we showed how the growth of a giant planet in the disk can have an impact on the accretion of solids onto the convective envelope of the disk. Therefore, we investigate whether the formation of a giant planet around one of the constituents of the HD106515 binary system is able to reproduce the observed differences in elemental abundances. For ease of comparison, similar to the considerations in Sect. \ref{sec:res_jup}, the present day abundances will be applied as the initial conditions of the system. In addition to the reason presented in that section, two simulations will be compared in this one, with and without the creation of the planet, such that only the difference between those two cases is relevant. Therefore, being off from the real initial condition can be neglected here for simplicity. Note that, while it seems possible to draw a straight line at $\Delta\mathrm{[X/H]}=0.001$ to explain the abundance differences within the measurement uncertainties (see Fig. \ref{fig:hd_summary_pl}), it might be questionable why, in this case, the measurement of oxygen should be that far off, meaning why it is mostly a negative abundance. We thus think that these measurements point to an interesting scenario for planet formation and future observations will show if these features can be recovered in other binary systems.

As the two stars are not identical, chemical abundances of star B are used for the cases with and without a giant planet as initial conditions, and small effects created by the change of these properties between the two stars are neglected. In particular, the properties of star B were chosen as opposed to star A because the effects of the formation of a planet in the disk are considered, which already influenced the abundances of star A. High precision measurements used to obtain the abundance differences do not give the individual stellar abundances \citep{liu2021}, hence the measurements of the individual stars by \citet{saffe2019} are used, with the abundance based on the Ti I line is applied for titanium. The mass of the central star in the simulations was chosen to be \SI{0.88}{\solarmass}. While the stellar model that is in the closest agreement with these parameters employs that mass, it uses $\mathrm{[Fe/H]=0}$. Like for previous considerations, the planet's location is kept fixed. The effect of planetary migration is shown in Appendix \ref{sec:res_hd_mig}. The employed approach assumes that all model parameters, apart from the presence of the planet, are identical for the accretion disks around both stars. We choose a disk radius of \SI{75}{\astronomicalunit} and disk mass of $0.1\ \mathrm{M_\star}$. This is done for simplicity, but in reality, the disks around the stars might have been different. A scenario where the two disks are of unequal initial mass is presented in Appendix \ref{sec:hd_mdiff}. Like discussed in Sect. \ref{sec:res_nopl}, a difference in disk parameters between the two disks can have an impact on the difference of the final convective zone abundances. However, the focus here lies on the impact of planet formation in otherwise identical disks.

First, the planetary seed in system A is placed at the present-day observed location. For different sets of grain size-altering parameters, including only those that allow growth to pebble isolation, the top of Fig. \ref{fig:hd_summary_pl} shows the resulting contrast between the convective zone abundances of the stars whose disks did and did not form a giant planet, respectively. It is apparent that in this case, the produced contrast is too large for the majority of elements, even in the case of the largest particles, where the impact of the forming giant planet is the smallest (see Sect. \ref{sec:res_pl}).

Furthermore, a key trend is found in the observations that is not reproduced with this setup: The abundance difference between stars B and A not only drops substantially compared to carbon and most refractories, but also likely negative. In fact, the present-day location of the planet corresponds to a location outside the water ice line, so that the planet blocks the accretion of water ice alongside the refractories. Therefore, the final stellar abundance of oxygen does not differ significantly from other elements in simulations where the planet is fixed at that location. The drop in abundance difference can, however, be reproduced in simulations if the planetary seed is placed such that the accretion of water ice is not affected by the pressure bump the planet creates. In that case, the difference of the oxygen abundance drops.

Reproducing $\Delta\mathrm{[O/H]}<0$ is not possible if planet formation, without migration, in one of the disks is the only considered effect, and the disks have the same initial mass. The negative difference implies that the star whose disk did not form a planet experienced the accretion of less enriched material than when a planet is formed, contradicting results of previous sections. In addition, the simulated abundance differences are too large even if the planet forms inside the water ice line.

A solution to these issues for disks of the same mass could lie in the formation of plantesimals, since that also prevents the accretion of solids from the disk by locking them those objects. While a realistic model of planet formation involves forming planetesimals in the disks of both binary constituents, the abundance differences found when their formation is not considered suggests a higher efficiency of locking-up material in the disk where no planet forms. Therefore, in the interest of simplicity, planetesimal formation is only included in the simulation of the disk around star B.

\begin{figure*}[htp]
    \centering\includegraphics[width=0.49\textwidth]{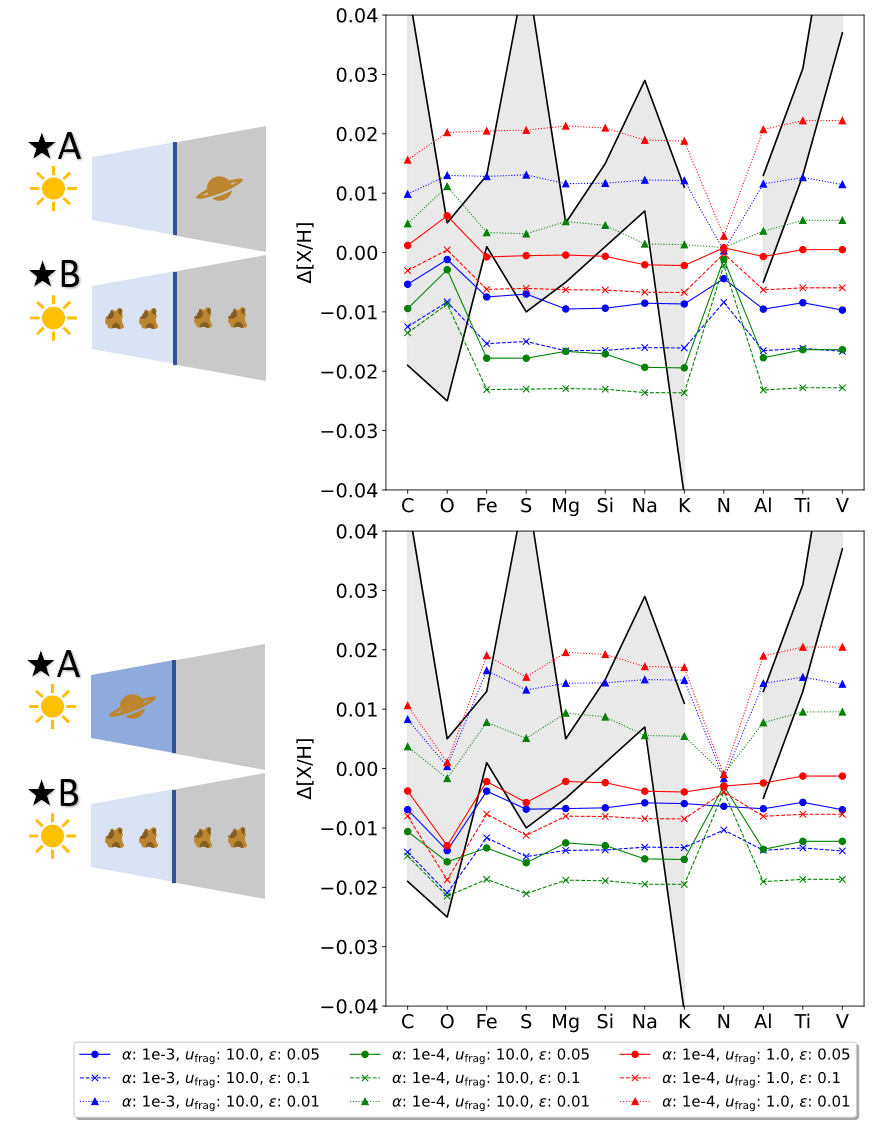}
    \centering\includegraphics[width=0.49\textwidth]{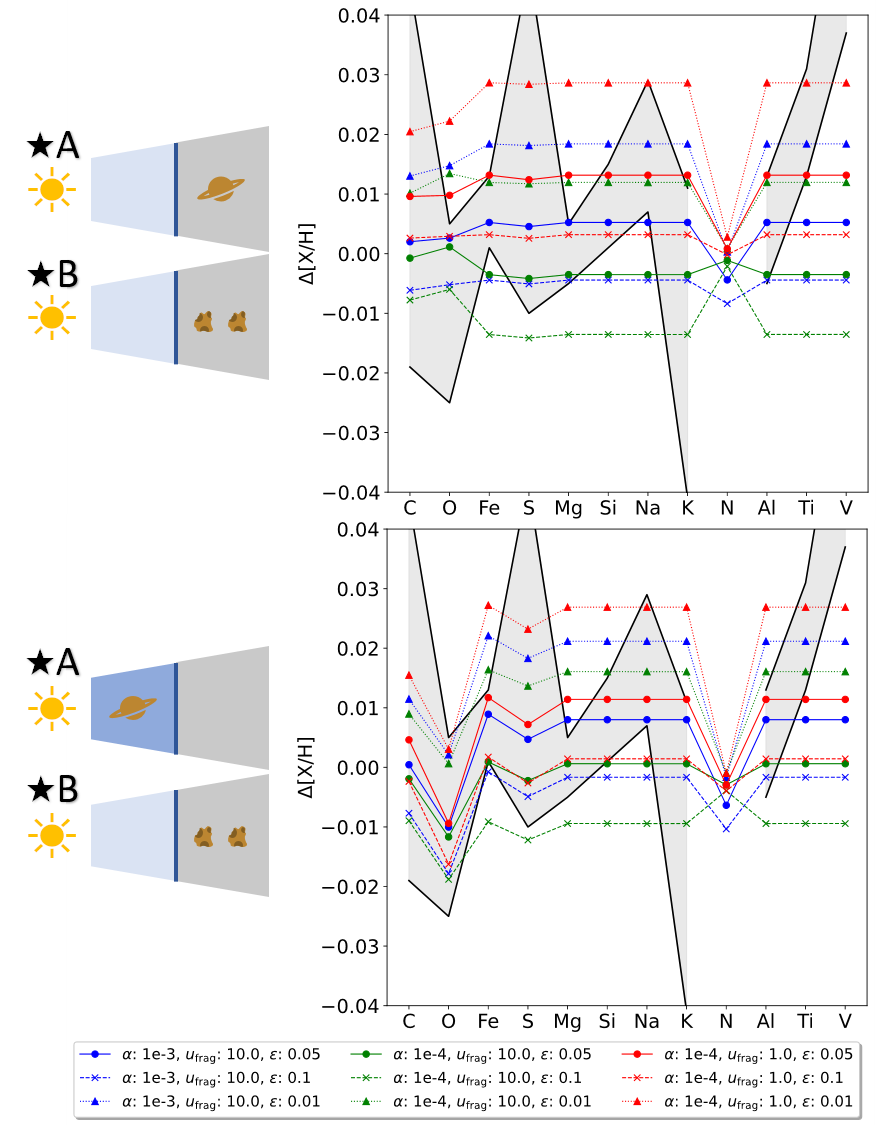}
    \caption{Same as Fig. \ref{fig:hd_summary_pl}, but including planetesimal formation around star B. For simulations shown on the left-hand side, plantesimals can form in any disk region where the pebble flux is above a threshold value, in accordance with the model by \citet{voelkel2020}. For those on the right-hand side, planetesimal formation is turned off for $r<r_\mathrm{H_2O}$. The planetesimal formation efficiency $\epsilon$ is varied between lines with different markers and styles. Sold lines with a circular marker show the results for $\epsilon=0.05$, dashed lines with a cross marker show $\epsilon=0.01$ and dotted lines with triangular markers show $\epsilon=0.1$.}
    \label{fig:hd_summary_both}
\end{figure*}%
Figure \ref{fig:hd_summary_both} presents the scenario of planetesimal formation in star B's disk, where three different planetesimal formation efficiencies $\epsilon\in\{0.01,0.05,0.1\}$ are considered. Two different scenarios are shown. In the first scenario, shown on the left-hand side of Fig. \ref{fig:hd_summary_both}, planetesimals can form anywhere in the disk given a high-enough pebble flux, while they cannot form interior to the water ice line in the second scenario, shown on the right-hand side of Fig. \ref{fig:hd_summary_both}. The second scenario is motivated by the planetesimal formation model by \citet{drazkowska2017}, where pebbles are too small to form planetesimals inside the water ice line. Note that this is caused by a radial change in fragmentation velocity in their model, whereas we employ a constant value. If planetesimals can form anywhere, but the pebble flux is low or planetesimal formation is inefficient, it predominantly occurs in the inner disk regions, where the dust composition shifts toward higher fractions of refractories. If planetesimal formation can only take place in the outer disk, where also volatiles are in solid form, all species are locked-up in equal fractions. Since planet formation also prohibits the accretion of refractories whose evaporation front locations are blocked equally, the second case shows a constant abundance difference per condensation temperature for these elements.

In simulations where star A's planet forms at the present-day location, the observed oxygen abundance trend is not reproduced, regardless of the planetesimal formation scenario. If planetesimal formation can occur anywhere in the disk, the resulting abundance difference is negative or close to zero for all elements at medium and high formation efficiency ($\epsilon\geq 0.05$) and large particles ($u_\mathrm{frag}=\SI{10}{\meter\per\second}$), not just for oxygen. In these simulations, locking-up material in planetesimals is more efficient at preventing enrichment of heavy elements than the planet's pressure bump. In fact, the case where the largest pebbles are formed benefits locking up material due to planetesimal formation the most compared to blocking by the planet. Here, larger particles are less efficiently blocked by a pressure bump (see Sect. \ref{sec:res_pl}), while a large mass is locked up in planetesimals due to the high pebble flux. The pressure bump of the planet, located outside the water ice line, blocks the flux of icy pebbles, and the low condensation temperature of water does not result in significant inclusion in planetesimals. Hence, the relative importance of the two effects is reversed for large particles, so that those simulations show an upward bump for $\Delta\mathrm{[O/H]}$, opposite to the observations. On the other hand, for the lowest efficiency $\epsilon=0.01$ and smaller particles, the influence of the planet is larger than that of planetesimal formation. If planetesimals do not form in the inner disk, their influence is also diminished compared to the pressure bump. Therefore, the abundance differences shift more toward positive final values. The difference in inclusion efficiency per species is reduced, as most species are solid in the outer disk. Here, only configurations of $\epsilon=0.1$ and $u_\mathrm{frag}=\SI{10}{\meter\per\second}$ reach a negative final difference. Although the contrast between the convective zone abundances of the disk with a planet and without agrees for some configurations with the measured $1\sigma$ interval regardless of whether planetesimal formation is limited to the outer disk region or not, it fails to reproduce the observed negative difference between stars B and A for oxygen while also producing a positive difference for the other elements.

Results where such a trend is reproduced can be found if the planet is placed inside the water ice line, not hindering the accretion of oxygen in water. If planetesimal formation is not restricted, the most favorable results are produced for $\epsilon =0.01$ and for the largest grains. Here, both locking up material by planetesimal formation and pressure bump trapping is not very efficient, so that the contrast between the two star's convective zone is close to zero, while the fact that the disk around star B forms planetesimals allows the contrast in [O/H] to be negative. Higher efficiencies lead to the other abundance difference values becoming negative as well. If planetesimal formation occurs only in the outer disk, a higher efficiency parameter is required to match observations. A medium efficiency of $\epsilon=0.05$ is found to be in good agreement with the measurements for the largest grains. The value for $\Delta\mathrm{[O/H]}$ is more negative in this case, while the differences for all other elements is positive. This is because restricting the planetesimal formation to the outer disk means that fewer refractories are locked up in planetesimals overall, moving their abundance difference toward positive values, while the high formation efficiency and pebble flux in the outer disk create a negative contrast for oxygen, not affected by the planet. Disks with $\epsilon=0.1$ and the smallest grain size can agree with the measured differences, too.

The simulations presented in Fig. \ref{fig:hd_summary_both} show disks with $M_0=0.1$ and $R_0=\SI{75}{\astronomicalunit}$. A reduced disk mass moves the final abundance difference closer to zero overall, allowing more configurations, mainly those featuring lower formation efficiencies, to match the observations. A larger disk radius of \SI{150}{\astronomicalunit} is not investigated here, because such large disks are close to overlapping, given the binary semi-major axis of $345^{+95}_{-47}\ \si{\astronomicalunit}$.

The results found in this section show that it is possible for planet formation to be the cause for the elemental abundance difference measured for the HD106515 wide binary system by influencing the material that is accreted onto the convective envelope of the central star. However, to reach appropriate levels for the differences, especially $\Delta\mathrm{[O/H]}<0$, additional requirements have to be met. One possibility is including the formation of planetesimals. In that case, the observations are reproduced only for cases where the disk that does not form a planet (around star B) forms planetesimals and the disk around star A does not, representing the physical case where planetesimal formation is more efficient in the disk that was unable to form a planet. This could possibly indicate that efficient planetesimal formation is not beneficial for giant planet formation. Alternatively, the disk that does not form a planet has to be lighter than the one that does (see Appendix \ref{sec:hd_mdiff}).

While the planet at its present-day location, outside the water ice line, can match the observations in some scenarios, the fact that the oxygen abundance difference drops significantly compared to the other elements cannot be reproduced this way. For that, the planet would need to be formed inside the water ice line. In this scenario, simulations with equal-mass disks and various disk and planetesimal formation parameters, as well as a simulation with disks of unequal mass, reproduce the observations for most elements.
\section{Discussion}\label{sec:discussion}
In this work, several simplifications were made to balance the required computational power and the level of physical detail the applied model can capture. Treatment of the evolution of the stellar convective zone was done by using stellar evolution models treating young stars in isolation. It was then assumed that the resulting evolution of the convective zone is applicable for calculating the change of abundances due to accretion of the disk. To compute that change, instantaneous mixing was assumed. Moreover, it was presumed that the convective envelope keeps its composition while transitioning away from a fully convective star. Regarding the composition, \citet{kunimoto2018} show that accurate abundances are reproduced under such simplifications. However, this approach falls short of considering that the accretion has an impact on the stellar evolution itself, hence changing the mass evolution of the convective envelope \citep{baraffe2010} and affecting how disk accretion is reflected on the elemental abundances. While \citet{kunimoto2017,kunimoto2018} show that accretion complicates the pre-main-sequence evolution of the star, the implication for this work is that the stellar interior is not necessarily fully convective even at early times, indicating that the here presented abundances differences could be higher, emphasizing the importance of connecting disc and stellar evolution.

\subsection{Grain size model}\label{sec:disc_grainsize}
The model used to compute the Stokes number of the dust in the disk, in turn setting the radial drift speed and the efficiency of interactions with an accreting planet, uses an approach treating only two distinct populations with representative sizes, using fits to find good agreement with simulations that treat the growth and fragmentation of dust in detail \citep{birnstiel2010}. A full treatment of dust is computationally involved (e.g., \citealt{stammler2022}) and is not feasible to be included in this model, which also considers planet formation and the evaporation and condensation of material. However, the downside of using a mass-averaged velocity is that differently-sized grains cannot separate in the disk. In reality, the small dust-grain population is not affected by a planet at pebble isolation strongly enough to block it \citep{ataiee2018,bitsch2018a,picogna2018}, allowing part of the solids outside the planet's orbit to pass it even at later times. \citet{stammler2023} find significant "leaking" of a planet-induced dust trap when grain fragmentation plays an important role, with 80\% of the dust outside the planetary orbit being accreted through the gap after \SI{10}{\mega\year} for a \SI{200}{\earthmass} planet, depending on parameter choices. This is not reproduced in this model, where it is only possible for the entire dust population to avoid being blocked if the resulting mass-averaged outward-pointing velocity is smaller than the inward-pointed gas drag. Furthermore, the 1D-approach allows the trapping of all pebbles if the planet is massive enough, which does not match a more realistic 3D scenario, where small pebbles could be transported through the planetary orbit via meridional flows enhanced by vertical shear (e.g., \citealt{picogna2018}). Altogether, the effects of pebble trapping by a massive planet would be diminished if such a more detailed treatment was included in the model.

\subsection{Disk temperature and mass loss}
The location of the ice lines in the disk play a crucial role for the strong enrichment of refractories early on in the simulations and the late accretion of material enriched in volatiles. For simplicity, the model used here employs a temperature profile constant in time. However, a more realistic scenario needs to treat changes of temperature by a change in the stellar luminosity, the surface density of the disk, the dust-to-gas-ratio change due to dust radial drift and radiative cooling (e.g., \citealt{bitsch2015a,savvidou2020,savvidou2021}). Including such condensations could shift the ice lines in the disk based on the evolutionary stage of the system. Ultimately, the details of the temperature structure of the disk can be relevant to find a realistic placement of the ice lines over time, especially relative to a massive planet in the disk. If the ice line is, e.g., outside a planet's orbit at first but moves inside it as the disk cools, a stellar abundance on a level between the case without any planet and one with full trapping can be achieved.

Stellar irradiation does not only heat the disk, but photoevaporation creates steady mass-loss. In the model used here, the mass-loss is only considered in the form of rapid depletion once the disk lifetime is reached. However, the dust surface density is not strongly affected by the removal of disk mass, as large grains settle toward the mid-plane. Only smaller particles coupled to the gas could be carried away by photoevaporation winds, making the dust mass loss negligible (e.g., \citealt{owen2011}). However, photoevaporation can carve a gap in the disk, thereby creating a pressure bump that can block pebbles (e.g., \citealt{ercolano2022}). The difference to a gap formed by a giant planet is that a gap carved by photoevaporation also prevents further gas accretion, preventing a further change of convective envelope abundance. Disk winds, caused by photoevaporation, hydrodynamical or magneto-hydrodynamical effects have been found to influence disk evolution greatly by allowing for strong angular momentum transport (e.g., \citealt{chambers2019,lesur2022}). While a full-on treatment of disk winds is computationally unfeasible for this work, a change in the accretion speed due to more efficient angular momentum transport without stronger turbulence would affect the timing of accretion, diminishing enrichment of the stellar envelope if it is still in a stage where it is massive. A disk-wind driven scenario would require little turbulence in the mid-plane, in turn allowing pebbles to grow to large sizes. If the disk winds are strong enough, outward migration is possible even for single giant planets \citep{kimmig2020}, giving rise to various migration pathways crossing ice lines in both directions. Furthermore, the treatment of magnetic fields would constitute a more complex magneto-hydrodynamical model and can affect the accretion of the ferromagnetic iron.

\subsection{Giant planet formation}
The timing of the formation of giant planets that are massive enough to have an impact on the stellar envelope abundances is an uncertainty with this approach. The planet needs to form early enough to have an impact by trapping remaining material in the disk, for which it needs to reach the pebble isolation mass quickly before most material is accreted onto the star. While this is typically a requirement for giant planets to form their massive core, the question remains about how early the formation occurs relative to start of the pre-main-sequence evolution, which in turn defines the impact disk accretion has by setting the mass of the convective zone. If planet formation generally occurs early-on during the pre-main-sequence evolution, the impact is hindered by a more massive convective envelope. On the other hand, for super-Earths, the efficiency of trapping material is likely small, as they might not form early enough and not grow massive enough to block a large pebble mass. As was shown in Sect. \ref{sec:res_nopl}, the disk lifetime is indicative for the final elemental abundances, too. Refractory abundances typically reach their maximum value within the first ${\sim}\SI{2}{\mega\year}$, while a longer lived disk allows for the enrichment of volatile material at the expense of diluting the refractory abundance.

A correlation between the occurrence rate of giant planets and the metallicity of stars around which they occur was observed previously \citep{johnson2010,santos2004,fischer2005}, where more metal rich stars are more likely to harbor giant planets. Since giant planets trap refractory pebbles, stars where they form experience less enrichment in iron by disk accretion than stars without giant planets. Therefore, the measured correlation could be stronger if the initial metallicity was considered instead of present-day values.

\subsection{Abundance difference trend with condensation temperature}
Observations of the HD106515 system show no clear trend of the elemental abundance difference with condensation temperature. However, such trends are found for other systems with no known planets \citep{liu2021} and commonly discussed in the context of the Solar refractory depletion (e.g., \citealt{melendez2009}). In this work, most of the observed elements with high condensation temperature are not included in the chemical model, and a trend with condensation temperature was not investigated. This is because the evaporation fronts of the refractory molecules lie close together and close to the star given the temperature profile at hand, so that no significant difference can be created, neither by a difference in accretion timing nor by planet-induced trapping. Furthermore, the species with the highest condensation temperature in the model is TiO (see Tab. \ref{tab:partitioning}), and the aluminum-containing Al$_2$O$_3$ has a higher condensation temperature than VO, which is the only molecule containing vanadium here. Observations show $\Delta\mathrm{[Al/H]}<\Delta\mathrm{[Ti/H]}<\Delta\mathrm{[V/H]}$. This contradicts the general condensation temperature trend and the ordering based on molecular condensation temperature we use here. Using this approach, those difference therefore cannot be reproduced accurately. Even in cases where the trend matches the order of the partitioning model, differences between refractories would require a hotter disk so that the evaporation fronts are farther out and have greater spacing, allowing accretion timing or a pressure bump to influence refractories with higher condensation temperature differently from those with a lower one.

Generally, the technique we applied to find signatures of planet formation around a star and obtain information about the formation location can be applied to systems other than HD106515, too. However, it is not possible to find signatures of planet formation around field stars this way, because a comparison to a star who did not form a planet in its disk and had the same initial abundance is necessary. GALAH+ data show that Galactic variations of stellar initial conditions are considerably stronger than differences induced by planet formation, making statistical approaches not viable. This is also reflected by the error bars in Fig. \ref{fig:feh_bins}. Wide binary systems are favorable, as close binaries exhibit interactions whose treatment would warrant for a more complex model of planet formation and disk evolution. Including the formation of multiple planets in the simulations would allow, for example, the investigation of the HD133131 system, where the one constituent harbors one, and the other harbors two planets, or the abundance pattern of TW Hya, where a multi-gap model has been proposed to explain observations \citep{mcclure2020}.

\section{Summary}\label{sec:summary}
In this work, we considered the growth and drift of millimeter-sized dust grains in protoplanetary disks, separated into chemical species using a partitioning model, to find the change of the elemental abundances in the shrinking convective envelope of the central star due to the accretion of disk material. In addition, we investigated the impact of a growing giant planet in the disk and the order of magnitude of the expected abundance change for the Sun. We now summarize our findings as follows:
\begin{enumerate}
    \item For large dust grains, the efficient radial drift creates a strong enrichment of refractories in the accretion steam at early times, in turn raising the respective elemental abundances. At that point in time, the convective zone of the star is still very massive, so that the accreted material's composition is imprinted only slowly onto it. After the pebble reservoir of the outer disk is depleted, the material left in the disk is depleted in refractory species. Now, the disk is accreted onto a shrinking envelope which adapts to the disk accretion flux, only enriched in volatiles, more quickly. Due to the absence of refractories at that stage, the late accretion increases the volatile abundance while simultaneously diluting the refractory enrichment from the early accretion phase. The relative significance of this process is directly related to the disk lifetime, where a disk that is short-lived leaves behind a star whose convective zone is more enriched in refractories and less enriched in volatiles than if the disk was long-lived.
    \item The abundance changes of the stellar convective zone caused by disk accretion can reach up to ${\sim}\num{5e-2}$ for refractories and volatiles, where the highest level enrichment of refractories is reached if particles in the disk are large, i.e. for low viscosity or high fragmentation velocity. For volatiles, the highest enrichment is reached if the disk viscosity is high. There are other disk and stellar parameters that affect the final abundance difference, too. Stellar mass and metallicity affect the mass evolution of the convective envelope, in turn changing the relative significance of early and late accretion. More massive disks can create larger enrichment when accreted. Smaller disks reduce the enrichment in most cases by reducing dust growth and drift timescales, shifting the period of enriched accretion toward times when the convective envelope is more massive and adaptation to the composition of the accreted material is slow. Overall, the elemental abundances of the stellar convective envelope after the disk has dissipated differ from the initial ones present in the natal molecular cloud, and in turn initially in the star and the disk.
    \item The formation of a massive planet changes the material accreted onto the convective envelope after reaching the pebble isolation mass by trapping solids outside its orbit. The effect of the growth of a planet in the disk is defined by the time it takes the initial protoplanetary seed to grow to this mass limit, and the amount of material that is still left outside the planet's orbit at that point. Moreover, the composition of the blocked solids is related to the position of the planet relative to the evaporation fronts. Molecular species with ice lines outside the planet's orbit are not affected by its presence, having already evaporated before they pass its location. Also, dust grains need to be of a certain size to be trapped, as the strong coupling to the gas and the resulting gas drag of small particles helps them overcome the pressure bump that is created outside the planet's location.
    \item As water is the most massive heavy component in the disk, the oxygen abundance in the convective envelope is particularly strongly affected by the formation of a planet outside or inside the water ice line, creating changes in the stellar envelope [O/H] between those two locations of ${\sim}\num{e-2}$. Given observations of this precision, the elemental abundances of stars can be used to constrain the formation location relative to ice lines of planets that formed in their nebula. Refractory material is affected by the formation of a planet in most cases, as their evaporation fronts are close to the star and likely to be inside the planets orbit in giant planet formation scenarios.
    \item The effect of disk accretion on the Sun's elemental abundances was investigated with the simplification of only considering Jupiter as the most significant gravitational influence in the disk. After \SI{4.5}{\mega\year}, the change in refractory abundances can reach ${\sim}\num{2e-2}$ for massive disks with high fragmentation velocity and low viscosity, but can be as low as ${\sim}\num{e-3}$ at $\alpha=\num{1e-3}$ with low fragmentation velocity and disk mass, and even lower at $\alpha>\num{1e-3}$. In contrast, such highly viscous disks can enrich the volatile nitrogen up to $\mathrm{[N/H]}\sim\num{1e-2}$. Depending on the parameters of the Solar protoplanetary disk, the initial abundances of the Sun therefore differ in an observationally discernible matter from today's values.
    \item Observations have revealed chemical abundance differences between stars in the HD106515 wide binary system. One of its constituents, HD106515A, harbors a massive giant planet at ${\sim}\SI{4.5}{\astronomicalunit}$, while the other has no confirmed planets. A key aspect is that $\Delta\mathrm{[O/H]}<0$, while all other $\Delta\mathrm{[X/H]}>0$. This is best reproduced if the planet, despite its present-day location, formed inside the water ice line, and if either planetesimals are formed more efficiently around star B, which does not harbor a planet, or the disk around star B was lighter. For the former, the requirement of more efficient planetesimal formation in disk B compared to A, where no planet has been observed as of today, might provide hints about the role planetesimals can play in the broader picture of planet formation. This approach shows that stellar abundances can give information about the formation history of planetary systems similar to what measurements of the atmosphere of the planets could provide, e.g., constrain the position of the planet relative to the water ice line or the efficiency of planetesimal formation. A combination of the constraints obtained by a combination of both types of observations in a single system could allow for further insights in the future.
\end{enumerate}
Our here presented model opens up new opportunities to constrain planet formation theories via stellar abundances and emphasizes the need for detailed stellar abundances of exoplanet host stars.

\begin{acknowledgements}
L.-A.H. acknowledges financial support from the European Research Council via the ERC Synergy Grant ECOGAL (grant 855130). B.B. thanks the European Research Council (ERC Starting Grant 757448-PAMDORA) for their financial support.
\end{acknowledgements}
\bibliography{references}

\begin{appendix}
\section{Dust drift model}\label{sec:dust_drift}
The \texttt{chemcomp} model, in addition, accounts for grain density ($\rho_s$) changes that arise as volatile and refractory elements condense and evaporate (see Sect. \ref{sec:meth_chem}) by dynamically calculating it during each time step following \citet{drazkowska2017},
\begin{equation}
    \rho_s = (m_\mathrm{ref}+m_\mathrm{vol})\left(\frac{m_\mathrm{ref}}{\rho_\mathrm{ref}}+\frac{m_\mathrm{vol}}{\rho_\mathrm{vol}}\right)^{-1}\text{,}
\end{equation}
with $\rho_\mathrm{ref}=\SI{3}{\gram\per\centi\meter}$ and $\rho_\mathrm{vol}=\SI{1}{\gram\per\centi\meter}$, as well as $m_\mathrm{ref}$ and $m_\mathrm{vol}$ the mass of refractory and volatile material in solid form, respectively. The grains are subject to radial drift, gas drag and turbulent mixing. The relative significance of these effects depends on the coupling strength of the dust to the gas, described by the Stokes number St. Additionally, the grains grow in size as they are subject to sticking collisions, increasing their Stokes number. As the gas particles move with sub-Keplerian speed due to the pressure gradient in the disk, the dust grains experience the headwind effect so that their radial drift speed exceeds that of the gas. It is given by
\begin{equation}
    u_Z=\frac{1}{1+\mathrm{St}^2}u_\mathrm{gas}-\frac{2}{\mathrm{St}^{-1}+\mathrm{St}}\Delta v\text{,}\label{eq:driftspeed}
\end{equation}
where $u_\mathrm{gas}$ is the radial velocity of the gas and $\Delta v$ is the difference of the azimuthal gas velocity to the Keplerian value. As long as $\mathrm{St}<1$, the difference to the gas velocity becomes very large as the particles increase in size. Grain growth is subject to size limits, with the dominant growth-limiting factors being fragmentation and radial drift \citep{dominik2008,brauer2008}.

Treating the former, the Stokes number at which relative velocities between similar-sized grains equal the fragmentation velocity $u_\mathrm{frag}$ sets an upper limit for the particle sizes \citep{birnstiel2009},
\begin{equation}
    \mathrm{St}_\mathrm{frag} = f_f\frac{1}{3}\frac{u_\mathrm{frag}^2}{\alpha c_s^2}\text{,}\label{eq:st_frag}
\end{equation}
where $\alpha$ is the turbulent viscosity parameter and $c_s$ is the sound speed. Lab experiments reveal $u_\mathrm{frag}$ values between ${\sim}\SI{1}{\meter\per\second}$ for silicates \citep{blum2008} and ${\sim}\SI{10}{\meter\per\second}$ for icy grains \citep{gundlach2015}. Both values will be considered in later sections, but a radial change of fragmentation velocity based on solid composition is not considered. Analytical models of the grain size distribution show that a significant fraction of the dust has sizes slightly below the limiting Stokes number in the fragmentation limited regime \citep{birnstiel2011}. Therefore, an additional model parameter $f_f$ is introduced in Eq. \ref{eq:st_frag}, which is set to $f_f=0.37$.

Furthermore, the size of particles at a given radial distance from the central star is limited by the drift of the particles. As they grow in size, their drift speed increases, making them drift away from their original location, thereby limiting the maximum particle size at that location. It is characterized by a maximum Stokes number,
\begin{equation}
    \mathrm{St}_\mathrm{drift} = f_d\epsilon_d\frac{v_K^2}{c_s^2}\gamma^{-1}\text{,}\label{eq:st_drift}
\end{equation}
with $\epsilon_d=\frac{\Sigma_Z}{\Sigma_\mathrm{gas}}$ the vertically integrated solid-to-gas mass ratio, $\Sigma_Z$ the solid surface density, $\gamma=\left|\frac{d\ln P}{d\ln r}\right|$ the absolute value of the power-law index of the gas pressure profile, and $v_K$ the Keplerian velocity. Analogously to Eq. \ref{eq:st_frag}, a model parameter $f_d=0.55$ is introduced to fit more detailed models.

The small population, being one of the two components of the model, is always characterized by the size of the smallest monomers ($a_0$), being subject to gas drag while exhibiting significant coupling to the gas. Lastly, to find the time and location dependent representative size of the large population, delayed drift has to be accounted for, as grains first have to grow to large enough sizes to be influenced by drift, growing slower in outer regions of the disk. The size of the large population is hence given by the minimum size set by the fragmentation, drift, and growth limit.

\section{Planet growth and pebble isolation model}\label{sec:meth_peb_iso}
The growth of one giant planet can be modeled using \texttt{chemcomp}, starting from a planetary seed and growing through pebble accretion (for a review, see \citealt{johansen2017}), which allows fast growth of massive planetary cores \citep{ormel2010,johansen2010,lambrechts2012,morbidelli2012}, up to the pebble isolation mass \citep{morbidelli2012,lambrechts2014,ataiee2018,bitsch2018b}, and subsequent envelope gas accretion (e.g., \citealt{ikoma2000,machida2010,ndugu2021}).

In the context of a planet's influence on the composition of material accreted onto the stellar convective envelope, the timescale of core growth by pebble accretion is responsible for allowing it to reach the pebble isolation mass. Therefore, it describes at what point during the disk's lifetime the planet creates a pressure bump strong enough to alter the dust evolution. Initially, the planetary seed is placed in the disk at the transition mass where pebble accretion becomes efficient \citep{lambrechts2012,johansen2017}. During the subsequent time evolution, accretion rates from \citet{johansen2017} are applied. They strongly depend on the Stokes number of the pebbles, and are the largest close to $\mathrm{St}=0.1$. To account for the evaporation of pebbles during accretion, which contributes material to the planetary gaseous envelope, only 90\% of the accreted pebble mass is used for the core growth, while the other 10\% are added to the envelope.

As the planetary core grows in mass, gravitational interactions with the disk lead to a partial gap being opened, which has an impact on the movement of dust grains \citep{paardekooper2006,rice2006}. If the core grows massive enough, the accretion of pebbles is halted, preventing grains from drifting to further-in regions of the disk and reaching the central star, and stops the core from growing further. The limiting mass for this effect to occur is the pebble-isolation mass. For its prescription, findings from \citet{bitsch2018a} are employed,
\begin{equation}
    M_\mathrm{iso} = 25f_\mathrm{fit}\si{\earthmass}\text{,}
\end{equation}
with
\begin{equation}
    f_\mathrm{fit} = \left[\frac{H_\mathrm{gas}/r}{0.05}\right]^3\left[0.34\left(\frac{-3}{\log(\alpha)}\right)^4+0.66\right]\text{.}
\end{equation}
Due to the pressure gradient changes caused by pile-up at the ice lines, the dependence of the pebble isolation mass on the pressure gradient was not included for simplicity. The formation of a pressure bump and subsequent opening of a gap due to gravitational interaction of the planet with the disk is modeled by a change in viscosity, where the gap profile is described as a Gaussian around the planet's location.

\section{Planet migration for HD106515A b}\label{sec:res_hd_mig}
\begin{figure}
    \centering\includegraphics[width=\linewidth]{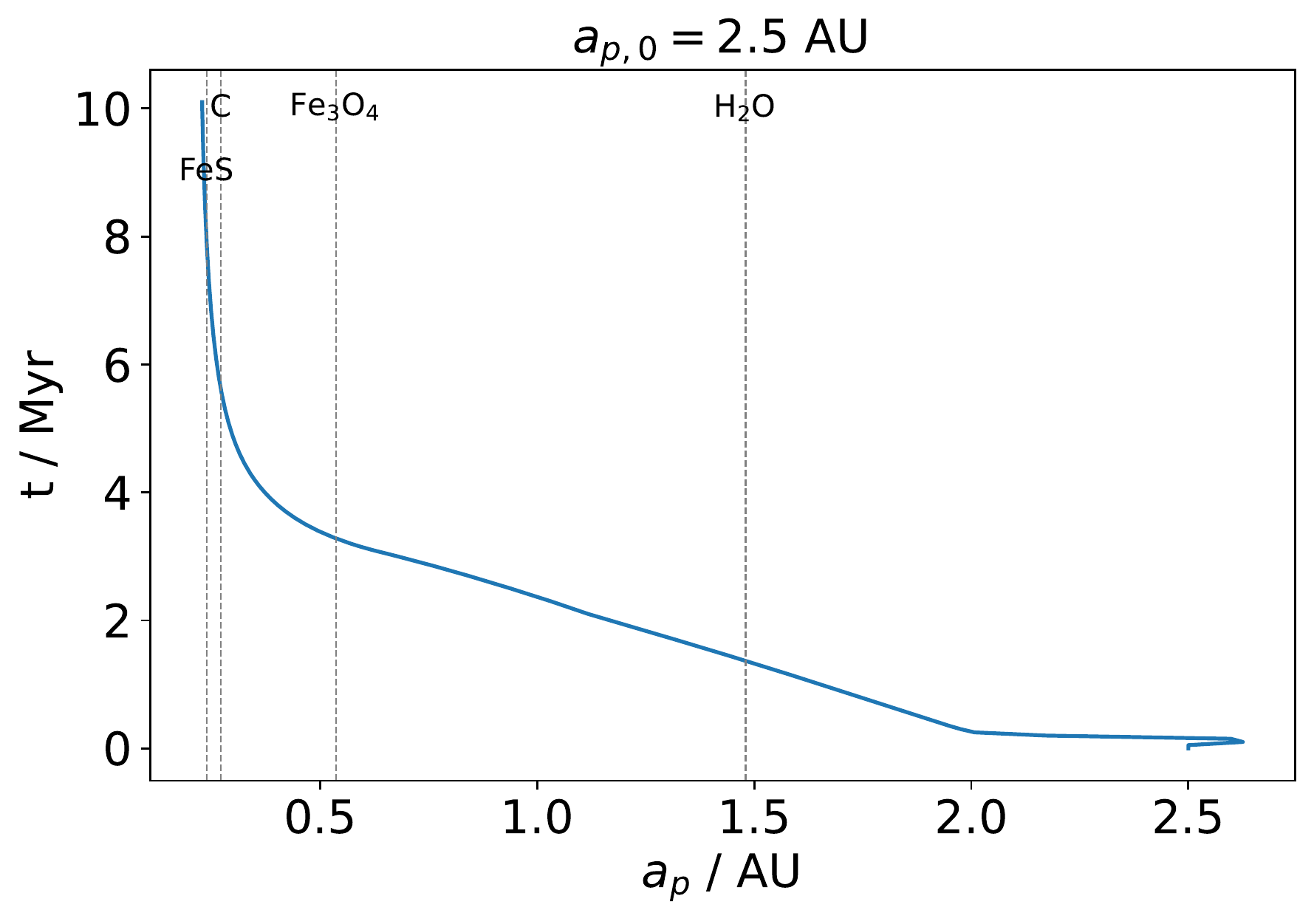}
    \caption{Migration track of HD106515A b for $a_\mathrm{p,0}=\SI{2.5}{\astronomicalunit}$, $\alpha=\num{1e-4}$ and $u_\mathrm{frag}=\SI{1}{\meter\per\second}$, shown in blue. The black dashed lines indicate ice lines that are crossed during the inward movement.}
    \label{fig:migration}
\end{figure}
\begin{figure}
    \centering\includegraphics[width=\linewidth]{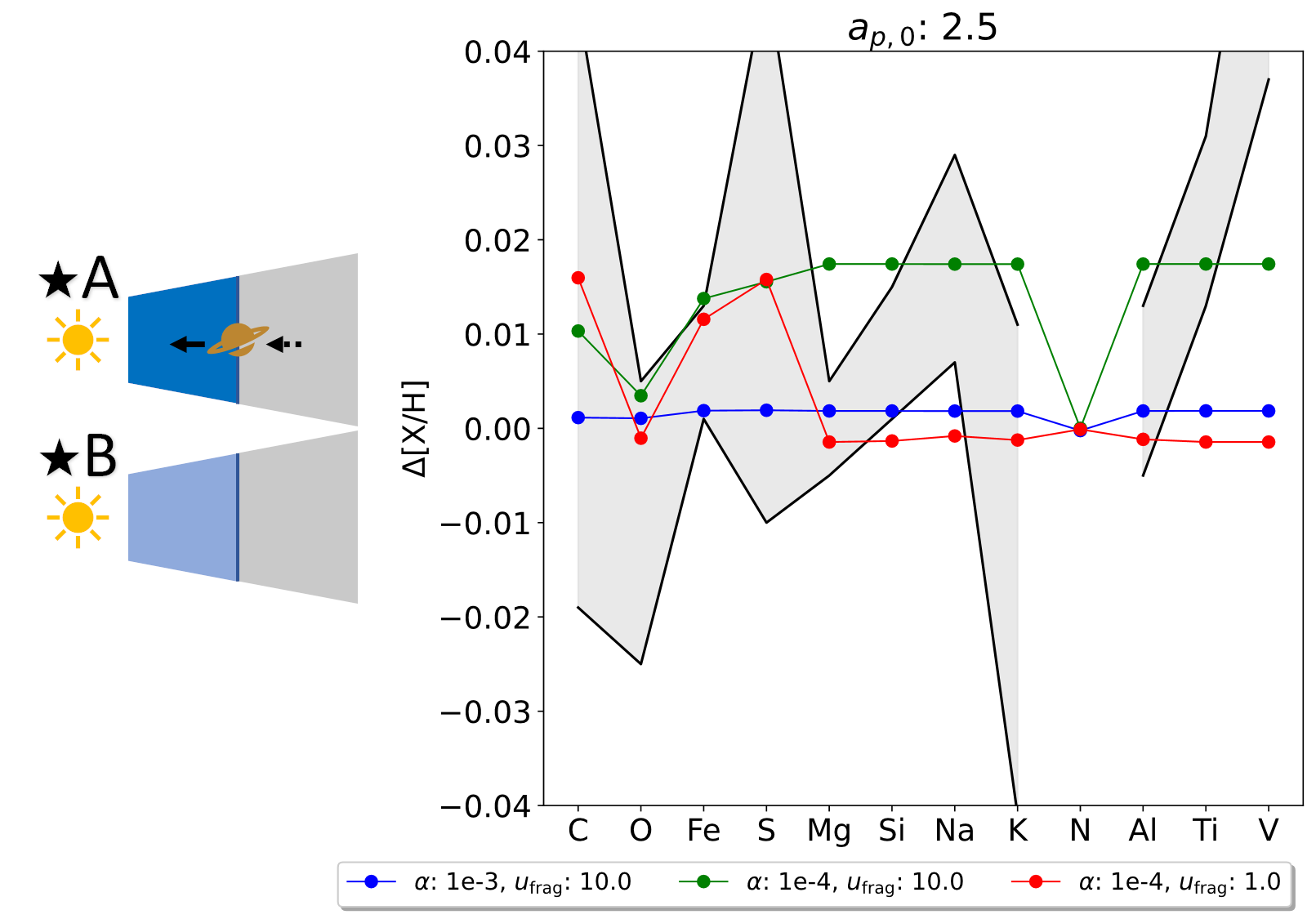}
    \caption{Like Fig. \ref{fig:hd_summary_pl}, but including migration of the planet around star A, with an initial position of $a_\mathrm{p,0}=\SI{2.5}{\astronomicalunit}$.}
    \label{fig:migration_hd}
\end{figure}
When treating planet formation around HD106515A without considering planetary migration, the negative abundance difference of oxygen between star B and A is reproduced only if star B, without any confirmed planets, forms planetesimals efficiently. In that scenario, it is possible for star A to accrete material more enriched in oxygen than star B, despite forming a planet in the disk. However, a second scenario is possible if the planet migrates. Here, a negative oxygen abundance can be created if the planet crosses the water ice line well into the disk lifetime. Initially, the position of the planet is chosen such that its pressure bump prevents the accretion of water. As it migrates inward, it eventually crosses the water ice line, allowing the icy pebbles that follow the planets migration path in the pressure bump to evaporate and be accreted onto the star. At this time, the stellar convective envelope has already shrunk, resulting in a faster adaptation of its composition to that of the accreted material.

In principle, due to the reduction in envelope mass, it is possible for the accretion of the previously blocked water to create an enrichment stronger than in a case where water was never blocked. However, this is very sensitive to the initial planetary location. If the ice line is crossed early, the blocked material does not give a significant contribution compared to a case where the initial position is inside the water ice line. In addition, the convective envelope is still close to its original mass at that time. On the other hand, if the ice line is crossed late, the oxygen-enriched gas is not accreted for a substantial amount of time before the disk dissipates. Therefore, even though the envelope has shrunk in mass at this time, a significant enrichment is not possible.

Due to this sensitivity to the initial position, only configurations with $\SI{2}{\astronomicalunit} \lesssim a_p \lesssim \SI{2.5}{\astronomicalunit}$ produce $\Delta\mathrm{[O/H]}<0$. In Fig. \ref{fig:migration}, the migration track of a planet with $a_\mathrm{p,0}=\SI{2.5}{\astronomicalunit}$ is shown, for $\alpha=\num{1e-4}$ and $u_\mathrm{frag}=\SI{1}{\meter\per\second}$. It crosses the water ice line at $t\sim\SI{2}{\mega\year}$, which is also the time when the convective envelope starts shrinking. At $t\sim\SI{4}{\mega\year}$, the Fe$_3$O$_4$ evaporation front is crossed, which allows a fraction of iron to be accreted onto a stellar envelope which is lighter due to its evolution.

In Fig. \ref{fig:migration_hd}, the final abundance difference are shown, including planetary migration. The best fit to the observations is produced by the $\alpha=\num{1e-4},u_\mathrm{frag}=\SI{1}{\meter\per\second}$. Like in the case of a fixed planet and planetesimal formation (see Sect. \ref{sec:res_hd}), individual trends for the highly refractory elements are not reproduced. In this migrating scenario, no substantial abundance differences are produced for those elements, because they are released and accreted onto the star at the end of the simulation, when the planet has reached the inner edge. In the case of $a_p=\SI{2}{\astronomicalunit}$, the highly refractory abundance differences even reach $\Delta\mathrm{[X/H]}\sim -0.01$, which contradicts observations. As the planet ends up at the inner edge of the disk, significant outward migration would be required to commence to find the planet at its present day location of \SI{4.5}{\astronomicalunit}. This problem is similar to the one of the Sect. \ref{sec:res_hd} scenario, where the planet needs to have formed inside the water ice line, corresponding to a formation inside \SI{4.5}{\astronomicalunit}. While in the former case, a solution could lie in the inward movement of the ice line due to disk cooling, outward movement of the ice line would be required to obtain results corresponding to the migration scenario while matching the observed planet location. Therefore, even though such outward movement can occur during the first few \SI{100}{\kilo\year} of disk evolution (e.g., \citealt{drazkowska2018}), the migration scenario might be less realistic.

\section{HD106515 model with different disk masses}\label{sec:hd_mdiff}
\begin{figure}
   \centering\includegraphics[width=\linewidth]{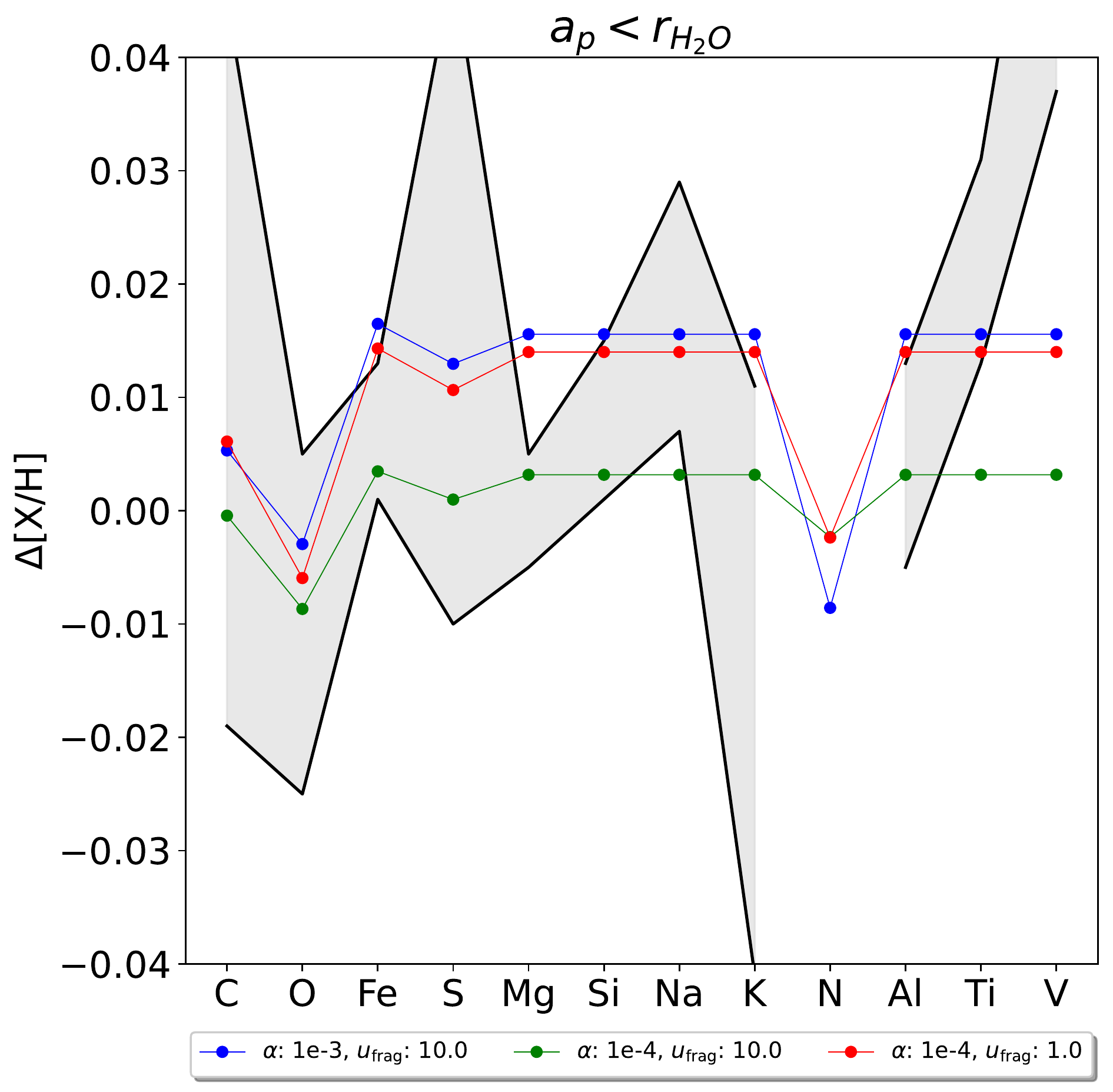} 
   \caption{Like Fig. \ref{fig:hd_summary_pl}, showing only the case where the planet formed inside the water ice line. Here, $M_{0,A}=0.1$ and $M_{0,B}=0.05$}
   \label{fig:mdiff_hd}
\end{figure}%
While the investigation of every possible variation of initial disk parameters between the binary constituents of HD106515 is outside the scope of this work, an interesting scenario is a difference in initial disk mass. The case where $M_{0,A}<M_{0,B}$ can be excluded, because the formation of a ${\sim}\SI{19}{\jupitermass}$ planet is difficult to explain in less massive disks with current formation models (e.g., \citealt{savvidou2023}). Additionally, reproducing $\Delta\mathrm{[O/H]}<0$ would be even more difficult in this case.

On the other hand, the case of $M_{0,A}>M_{0,B}$ is similar to the scenario where the disk around star B formed planetesimals more efficiently than the disk around star A. Instead of planetesimal formation removing material from the disk (see Fig. \ref{fig:hd_summary_both}), it has less mass initially. The main difference is that a change of $M_0$ affects the abundance of all chemical species equally, while planetesimal formation only affects materials that are predominantly in a solid phase. Because most species are in a solid phase in the outer disk, the described scenario is therefore most comparable to the bottom right scenario of Fig. \ref{fig:hd_summary_both}. The resulting abundance differences are shown in Fig. \ref{fig:mdiff_hd}. The best fit to the observations is produced when using the disk parameters $\alpha=\num{e-4}$, $u_\mathrm{frag}=\SI{10}{\meter\per\second}$. In particular, $\Delta\mathrm{[O/H]}<0$ is reproduced, and the abundance differences are within the measurement errors except for Na, Ti and V. In principle, further lowering $M_{0,B}$ could allow simulations with higher viscosities or lower fragmentation velocities to match the observations as well. However, further investigation would increase the parameters space, which is outside the scope of this work and would not change the conclusion that a lower disk mass around constituent B could help explain the observed abundance differences. 
\end{appendix}

\end{document}